%% file: main.tex
\documentclass[twocolumn,twocolappendix]{openjournal}

\PassOptionsToPackage{colorlinks,citecolor=blue}{hyperref}
\usepackage{amsmath,amstext}
\usepackage[T1]{fontenc}
\usepackage[utf8]{inputenc}
\usepackage[flushleft]{threeparttable}
\usepackage{url}
\usepackage{xspace}

\usepackage{orcidlink}

\usepackage[figure,figure*]{hypcap}

\usepackage{lsstdesc_macros}
\usepackage{graphicx}
\usepackage{url}
\usepackage{natbib}
\usepackage{tikz}
\usepackage{amssymb}
\usepackage{cases}
\usepackage[normalem]{ulem}
\usepackage{xcolor}
\graphicspath{{./}{./figures/}}

\usepackage{comment}
\usepackage{multirow}

\input{diagram.tex}

\newcommand{\ncode}[1]{\texttt{\bf#1}}

\newcommand{\clmm}{\code{CLMM}}

\newcommand{\ccl}{\code{CCL}}
\newcommand{\nc}{\code{NumCosmo}}

\newcommand{\qp}{\code{qp}}

\newcommand{\collab}[1]{\textsc{#1}}
\newcommand{\lsst}{\collab{LSST}}
\newcommand{\desc}{\collab{LSST-DESC}}

\newcommand{\paperI}{{Paper-I}}

\newcommand{\be}{back-end}

\newcommand{\mean}[1]{
    \langle#1\rangle
}

\newcommand{\rdmp}{\textsc{redMaPPer}~}
\newcommand{\metacal}{\textsc{metacalibration}~}

\newcommand\redout{\bgroup\markoverwith
{\textcolor{red}{\rule[.5ex]{2pt}{0.4pt}}}\ULon}

\newcommand{\SigCritEff}{\Sigma^{\rm eff}_{{\rm crit}\;ls}}

\begin{document}
\title{Getting ready for analysing Rubin data :\\
\clmm v2, the updated LSST-DESC Cluster weak Lensing Mass Modeling library for cosmology
}

\input{authors}

\begin{abstract}
The Dark Energy Science Collaboration has prepared a multitude of pipelines to process the upcoming Vera C. Rubin Observatory data and perform different scientific analyses. Here, we present the methodology to constrain the mass information of galaxy clusters from weak lensing, which is encompassed by the v2.0 release of the Cluster Lensing Mass Modeling (CLMM) code. This Python library provides the ingredients for both the modeling and data reduction of lensing observables, and the improvements in the new version were customized to handle the first WL analyses of LSST clusters, including, e.g., stacking functionalities and WL multipoles. This new version of CLMM has allowed various cluster analyses using LSST-DESC simulated data, precursor survey and Rubin commissioning and survey validation data, and is ready to be used for the first cluster cosmology analyses. CLMM is an open-source software package collaboratively written in Python, providing multiple example notebooks, including a set that can reproduce all figures in this work, and extensive documentation, publicly available at \url{https://lsstdesc.org/CLMM/}.
\end{abstract}

\begin{keywords}
    {galaxies: clusters: general; gravitational lensing: weak; software: public release}
\end{keywords}

\maketitle

\section{Introduction}
\label{sec:intro}

The Legacy Survey of Space and Time (LSST), started in late June 2026, is a wide-field imaging survey that will cover roughly half of the sky for ten years. It is operated by the Vera C. Rubin Observatory in Cerro Pachón, Chile, which features an 8.4-meter optical telescope equipped with a 3.2-gigapixel camera. The LSST will deliver some of the deepest and highest-quality optical imaging to date of the southern sky, enabling significant progress across multiple areas of astronomy. The survey will enable the scientific community to conduct detailed investigations of small bodies in the solar system, generate an extensive three-dimensional map of the Milky Way, uncover new insights into transient events, and characterize the properties of billions of galaxies across half of the sky \citep{lsst2012}.

Among the science collaborations that will exploit \lsst\ data, the Dark Energy Science Collaboration (\desc) focuses on understanding the nature of dark energy and dark matter using the main cosmological probes at visible wavelengths: cosmic shear and galaxy clustering, along with type Ia supernovae, strong gravitational lensing, and galaxy clusters \citep{2018arXiv180901669T}.

Galaxy clusters form through the collapse of large matter over-densities and emerge as today's largest gravitationally bound structures in the Universe. Their formation history, and their mass and redshift distributions are intimately connected to the amplitudes of fluctuations in the dark matter density field as well as the late-time expansion history of the Universe \citep[see e.g.][]{allenetal11}.

They are therefore a powerful cosmological probe to constrain the amount of dark energy and dark matter in the Universe, and can provide one of the strongest constraints on the growth of structure \citep{dodelson16}. This will be especially relevant with \lsst\ observations, from which hundreds of thousands of galaxy clusters are expected to be detected. This will increase by one to two orders of magnitude the size of the sample compared to that of previous cosmological analyses using optical survey data e.g., the SDSS\footnote{Sloan Digital Sky Survey} \citep{Sunayama2023HSClensing,Park2023lensingabundance,Fumagalli2024sdss}, the DES\footnote{Dark Energy Survey}\citep{desclustercosmo20,To2021DEScomb,Costanzi2021DESSPTCL,DES2025clY3}, the HSC\footnote{Hyper Suprime Cam} \citep{Chiu2024hsccl} or the KiDS\footnote{Kilo Degree Survey} \citep{Lesci2022KIDSCL}.

One of the key elements in cluster cosmology is the robust estimation of cluster masses \citep[see e.g.][for a review]{2019SSRv..215...25P}.  Accurate mass measurements are critical for calibrating the mass–observable relation, which links the observed abundance of galaxy clusters to cosmological models. In the past decades, weak gravitational lensing (in particular shear) has emerged as a powerful approach to do so: by measuring the subtle distortions in the shapes of background galaxies caused by the gravitational field of clusters, we can directly probe their total mass, without assumptions about their dynamical state.
With the advent of wide and deep Stage-III imaging surveys such as DES, KiDS and HSC, and in anticipation of next-generation Stage-IV surveys like the LSST and \textit{Euclid} \citep{Euclid2025} which will deliver lensing datasets to unprecedented depth and size, weak lensing has thus become a state-of-the-art technique for calibrating the mass–observable relation \citep{Simet2017SDSSmassrichness, Murata2018lensing, 2019PASJ...71..107M,McClintock2019mass,  Bocquet2024spt}.

    In this context, the DESC Cluster weak Lensing Mass Modeling library (\clmm)\footnote{\url{https://github.com/LSSTDESC/clmm}} has been developed over the past few years as an open-source Python package providing the ingredients necessary to compute cluster weak-lensing data vectors and predictions. The first version of the code, presented in \citet{Aguena2021clmmv1} (\paperI\ hereafter), offered tools for modeling idealized individual clusters, serving as a solid foundation for algorithmic development and validation. However, it was not designed to handle several systematic effects (miscentering, sample contamination, etc.) that the LSST real-data cluster applications will face. Furthermore, although we are still defining which specific cluster-lensing analyses will be undertaken once the first data release of LSST (DR1) becomes available (expected early 2028), a key missing capability in v1.0 was the ability to perform stacked lensing measurements that have been widely used in recent optical cluster cosmology studies \citep[e.g.][]{desclustercosmo20,Lesci2022KIDSCL,Sunayama2023HSClensing} and are likely to play a central role in early mass-calibration work. Beyond stacking, upcoming science cases could also require greater modeling flexibility, for example, the use of shear multipoles to characterize non-spherical mass distributions, motivating an expansion of the library functionality in preparation for real-data applications.

    These new capabilities have now been implemented in \clmm and were developed in close synergy with ongoing \desc\ science applications, including cluster-lensing analyses in the \desc\ DC2 simulated dataset (\citealt{Kovacs2022cosmoDC2,Payerne2025dc2CL,2026arXiv260912158R}), as well as studies of non-spherical cluster halos using weak-lensing multipoles \citep{2026arXiv260506587F}, the first WL analysis of a cluster observed with Rubin during comissioning (von der Linden et al. in prep.), and initial end-to-end cluster cosmology pipeline validation (Barroso et al. in prep.). Together, these developments represent a substantial evolution of the library and thus motivate the present v2.0 release, designed to support the first weak-lensing analyses of LSST clusters with DR1.

\clmm is part of a highly collaborative effort in DESC to develop public software for the analysis of galaxy cluster weak gravitational lensing for LSST. These codes come with extensive documentation and examples for the users, and most of them also have continuous integration for thorough testing implemented. As we move towards the first DESC cluster cosmological analyses, \clmm is now integrated in a larger analysis pipeline for a joint cluster abundance and cluster weak lensing analysis, that will also involve other DESC software such as \code{CCL}\footnote{\url{https://github.com/LSSTDESC/CCL}} \citep{2019ApJS..242....2C}, \code{TXPipe}\footnote{\url{https://github.com/LSSTDESC/TXPipe}} \citep{2023OJAp....6E..13P}, \code{sacc}\footnote{\url{https://github.com/LSSTDESC/sacc}}, \code{qp}\footnote{\url{https://github.com/LSSTDESC/qp}} \citep{2019AAS...23331305M}, \code{ClEvaR}\footnote{\url{https://github.com/LSSTDESC/clevar}}, \code{CROW}\footnote{\url{https://github.com/LSSTDESC/crow}}, \code{TJPCov}\footnote{\url{https://github.com/LSSTDESC/TJPcov}} and \code{fireCrown}\footnote{\url{https://github.com/LSSTDESC/firecrown}} {(see Appendix \ref{sec:interface} for an overview of the cluster LSST DESC pipeline architecture and dependencies).

This paper is organized as follows:
in Section \ref{sec:clmm},
we give a high-level description of the structure and changes on the new \clmm version, along with a brief overview of the v1.0 functionalities and limitations;
Section \ref{sec:new} details the new functionalities available on the theory and modeling side; in Section \ref{sec:new_data} we describe updates on the measurements and data handling part; and
in Section \ref{sec:application_on_data}, we illustrate some of the new functionalities using the code on observed or simulated datasets.
We provide a summary in Section \ref{sec:summary}.

For all computations and figures presented in this paper, we assumed the same cosmological parameters as in DC2: a flat $\Lambda$CDM model with $H_0$=71.0, $\Omega_{m0}$=0.265, $\Omega_{b0}$=0.0448 and $\Omega_{k0}$=0.0.
Additionally, unless stated otherwise, we assume a reference halo modeled with an NFW profile, at a redshift of $z=0.4$ and with a mass and concentration $M_{200,m}=10^{14}M\odot$ and $c_{200,m}=4$ (defined at a spherical overdensity $\Delta=200$ and with respect to the mean density of the Universe).
All figures presented in this work can be reproduced with the notebooks provided in the \clmm\ repository at
\hyperlink{https://github.com/LSSTDESC/CLMM/tree/main/examples/Paper_v2.0}{examples/Paper\_v2.0/}

\section{Overview of the code and of the main changes since v1.0}
\label{sec:clmm}

The main idea behind the development of \clmm is to provide a general cluster weak lensing software package to be integrated into the larger DESC cosmological pipeline, but that can also serve as a standalone library to be used more widely by the community. As such, the code development comes hand in hand with detailed documentation and a large tutorial notebook suite to ensure ease of use. Each functionality is also thoroughly tested, and the tests are implemented as part of continuous integration using GitHub Actions.

\begin{figure*}
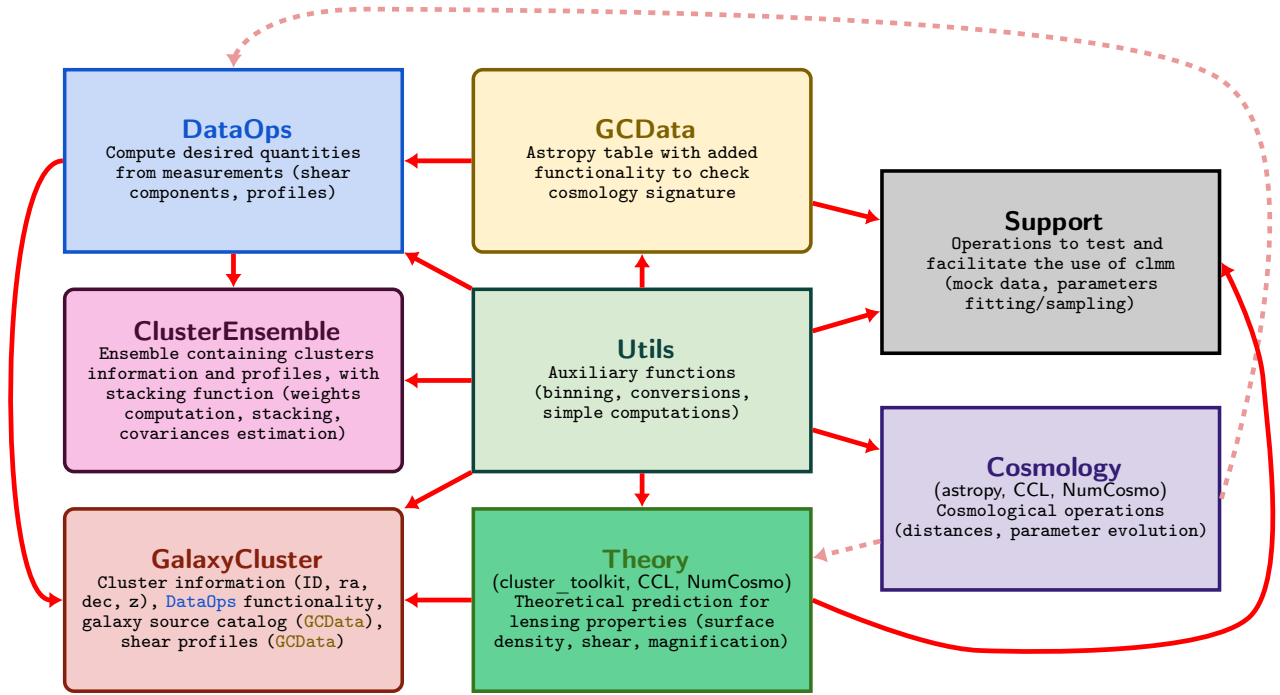

    \centering
    \resizebox{1\textwidth}{!}{\diagram}
    \caption{Diagram illustrating the structure of the \clmm code with dependencies (arrows).
    Sharp-edge rectangles represent packages, rounded-edge rectangles are modules, continuous arrows are direct imports, and dashed arrows mean the module is passed as an argument.
    }
    \label{fig:code_overview}
\end{figure*}

\subsection{Code structure}

Figure~\ref{fig:code_overview} summarizes the main modules and packages in this release, the structure of which remains mostly unchanged since v1.0. In this diagram, the main update is the new \code{ClusterEnsemble} module, created in v2.0 to enable data operations on an ensemble of clusters to ease (among other things) stacked WL analyses. The other modules have also improved, with many new functions or functionalities added in v2.0.
The core modules of \clmm are described as follows:
\begin{itemize}
    \item The \code{dataops} module deals with all computations to be performed at the data level, to go from background galaxy catalogs to data vectors, namely individual or stacked WL profiles, using either reduced shear or excess surface density. It relies on \code{GCData} (a modified version of an \code{astropy} Table) that not only stores the galaxy catalog but also does the bookkeeping of additional important metadata, such as the cosmology used in its processing.
    \item The \code{theory} module provides the corresponding theoretical predictions, given a cosmology (controlled by the \code{cosmology} module) and galaxy cluster properties.
    In v2.0, as it was in v1.0, all functions are still supported in both the functional interface and the object-oriented approach.
    \item The \code{GalaxyCluster} class provides a flexible object-oriented approach to apply \code{dataops} functionality to cluster data.
    \item The \code{ClusterEnsemble} module allows data operations on a list of \code{GalaxyCluster} objects, such as stacked radial profile measurement and data-driven profile covariance estimation.
\end{itemize}
These elements are at the core of \clmm and are supplemented by the \code{utils} package that provides low-level operations used in several places of the code base. The \code{support} package provides the user with additional features to perform end-to-end WL cluster analysis, as well as a mock data generator that we have widely used to build the suite of demo notebooks and tutorials.

\subsection{Dependencies and installation}

\clmm relies on various third-party software, listed in the documentation. In this release of the code the user can choose between \code{CCL} \citep{2019ApJS..242....2C} and \code{NumCosmo} \citep{2014ascl.soft08013D} for the theory back end (\code{cluster-toolkit}, \citealt{2022ascl.soft09004M}, is also available for functionalities of  v1.0, but it is not supported by new implementations), with \code{CCL} being the default choice if not specified otherwise by the user. A notable new dependency is the \code{qp} library \citep{2019AAS...23331305M}, which has been developed in DESC to provide a flexible framework to store and operate on probability density functions (PDFs). This library is used to handle redshift information, as it allows \clmm to provide a transparent way to work with either a point source estimation, a tabulated PDF on a regular parameter grid, or a quantile parameterization. Finally, the last new dependency is \code{healpy} \citep{2024zndo..12746571Z}, which is used to split the sky into various sub-regions before performing a Jackknife estimation of the stacked lensing profile covariance (see Section \ref{subsec:cov} for details)

For installation, the manual/source installation procedure already used in v1.0 still exists for those wanting to contribute to the code development. For non-development usage of the code, \clmm v2.0 is now also conda- and pip-installable to streamline the procedure further.

\subsection{Overview of the new functionalities}
On top of the new \code{ClusterEnsemble} module, \clmm v2.0 also brings many new functionalities in several of the modules and packages already defined in v1.0. These will be covered in detail in the paper.

To get an overall sense of what this new version brings,
    Table~\ref{tab:theory} provides a synthetic view of the main functions related to the theoretical
    prediction of the signal (top part) and to the measurements of the data vectors  (bottom part). For the functions that already existed in v1.0, we are only flagging the new functionalities that come with this version of the code and refer the reader to {\paperI} for more details on the original implementation. The new functions in v2.0 are highlighted in bold font.
The last column in the tables provides direct links to the sections where these new functionalities or new functions are discussed.

\begin{table*}
\centering
\caption{
Main theory/modeling-related and data functions implemented in \clmm. Functions in normal font have been updated with extended functionalities since the first release. Functions in {\bf bold font} are new functions added in v2.0. The last column gives the section of the paper where the new functionalities of existing functions, or the formalism for the new functions, are described.\label{tab:theory}}
\resizebox{0.95\textwidth}{!}{
\begin{tabular}{ l l p{1.7in} p{1.7in} c}
\hline \hline
\vspace{2pt}
Function name & Symbol & Output (prediction of...) & Update since v1.0 or new function & Section\\
\hline
\multicolumn{5}{c}{theory/modeling-related functions (* reads as \code{compute}/\code{eval} for the use of the functional/object-oriented interface)}\\
\hline \hline
\vspace{2pt}

 \code{*\_3d\_density} \dotfill &$\rho$ & Mass 3D density (NFW, Einasto or Hernquist)  & Free slope for the Einasto profile & --- \\
\vspace{2pt}

\code{*\_surface\_density}  \dotfill &$\Sigma$ & Surface density & Free slope for the Einasto profile, miscentering functionality & \ref{subsec:miscenter}\\
\vspace{2pt}

\code{*\_excess\_surface\_density}  \dotfill &$\Delta\Sigma$& Excess surface density & Free slope for the Einasto profile, miscentering functionality & \ref{subsec:miscenter} \\
\vspace{2pt}

\code{*\_convergence}  \dotfill & $\kappa$ & Convergence & Flexibility in source redshift & \ref{subsec:redshift}\\
\vspace{2pt}

\code{*\_tangential\_shear}  \dotfill & $\gamma_t$ & Tangential shear & Flexibility in source redshift, approximation when incomplete redshift information& \ref{subsec:redshift}, \ref{subsec:approx}\\
\vspace{2pt}

\code{*\_reduced\_tangential\_shear}  \dotfill & $g_t$ & Reduced tangential shear & Flexibility in source redshift, approximation when incomplete redshift information& \ref{subsec:redshift}, \ref{subsec:approx}\\
\vspace{2pt}

\code{*\_magnification}  \dotfill & $\mu$ & Magnification & Flexibility in source redshift, approximation when incomplete redshift information & \ref{subsec:redshift}, \ref{subsec:approx}\\
\vspace{2pt}

\ncode{*\_critical\_surface\_density\_eff}  \dotfill  & $\Sigma_{\rm crit}^{\rm eff}$  & "Effective" critical surface density computed from the galaxy redshift PDFs; used to compute the galaxy weights & New & \ref{subsec:weights}\\
\vspace{2pt}

\ncode{*\_magnification\_bias}   \dotfill & $\mu^{\alpha - 1}$ & Magnification bias & New & \ref{subsec:mag_bias}, \ref{subsec:approx}\\
\vspace{2pt}

\ncode{*\_excess\_surface\_density\_2h} \dotfill & $\Delta\Sigma_{2h}$  & 2-halo term excess surface density, Equation~\eqref{eq:2h_term} & New & \ref{subsec:2h}\\
\vspace{2pt}

\ncode{*\_excess\_surface\_density\_triaxial}   \dotfill & $\Sigma^{\rm const}, \Sigma^{4\theta}$ & Quadrupole components of the excess surface density & New & \ref{subsec:triaxial_mod}\\
\vspace{2pt}

\ncode{eval\_rdelta}   \dotfill & $r_\Delta$ & Overdensity radius, defined by $\Delta$ & New & \ref{subsec:mass_conversion}\\
\vspace{2pt}

\ncode{eval\_mass\_in\_radius}   \dotfill & $M(<r)$ & Mass inside 3D-radius $r$ & New & \ref{subsec:mass_conversion}\\
\vspace{2pt}

\ncode{convert\_mass\_concentration}   \dotfill & --- & Conversion between mass definitions & New & \ref{subsec:mass_conversion}\\

\ncode{compute\_\{nfw,powerlaw\}\_boost}\dotfill & $B$ & Radial boost model following a NFW or power-law $\Delta\Sigma$ parameterisation & New & \ref{subsec:boost} \\
\hline

\hline
\multicolumn{5}{c}{
data functions (for \code{dataops}, \code{galaxycluster}, or \code{clusterensemble}.
modules)}\\
\hline
\vspace{2pt}

 \code{compute\_tangential\_and\_cross\_components}\dotfill & --- & Angular separation to the source center and the tangential and cross values, given shape parameters. Also returns quadrupolar shear components if orientation information is available  & Option to choose coordinate system; option to compute the quadrupole shear components as well. & \ref{subsec:coord}, \ref{subsec:triaxial_meas}\\

\vspace{2pt}
 \code{make\_radial\_profile}\dotfill & --- & Radial profile of the tangential and cross components and associated errors in each bin    & Uses lensing weights in the averaging in radial bins & \ref{subsec:weights}\\

 \vspace{2pt}
 \ncode{compute\_galaxy\_weights}\dotfill & --- & Individual lens-source pair weights, from shape measurement errors and redshifts & New & \ref{subsec:clusterensemble}\\

\vspace{2pt}
  \ncode{make\_individual\_radial\_profile}\dotfill & --- & Same as \code{make\_radial\_profile} but method of \code{ClusterEnsemble} that stores the output in the data attribute of \code{ClusterEnsemble} object & New & \ref{subsec:clusterensemble} \\

\vspace{2pt}
  \ncode{make\_stacked\_radial\_profile}\dotfill & --- & Weighted average of individual radial profiles in a \code{ClusterEnsemble}  & New & \ref{subsec:clusterensemble}\\

\vspace{2pt}
\ncode{compute\_*\_covariance}\dotfill & --- & Covariance of a \code{ClusterEnsemble} stacked profile measured from the data, where \code{*} is \code{sample, bootstrap} or \code{jackknife} & New & \ref{subsec:clusterensemble}\\

  \ncode{compute\_background\_probability}\dotfill & --- & Probability of a galaxy to be behind the cluster, based on its redshift pdf   & New & \ref{subsec:backprob} \\

 \hline
\end{tabular}}
\end{table*}

\section{New theory and modeling functionalities}
\label{sec:new}

The new version of \clmm\ preserves the same structure as before, with the different functionalities split into modules, as can be seen in Fig.~\ref{fig:code_overview}.
In this section, we present the new theoretical modeling functionalities and improvements of \clmm\ v2.0 to increase the realism of the modeling.

Ongoing development of the theory subpackage, both directly in \clmm\ or its \be s, aims to improve the prediction of the weak lensing signal.
We anticipate carrying out informative cross-checks between multiple implementations of these models and sharing those results with the community.

\subsection{Two-halo term}
\label{subsec:2h}
In the halo paradigm \citep[a formalism that describes the spatial statistics of the dark matter density field, see e.g.][]{2002PhR...372....1C} for cluster lensing, the lensing signal produced by the mass distribution of the cluster itself is referred to as the 1-halo term. This contribution was at the core of the development of the theory module in \clmm v1.0, with the option to use various profiles (e.g., NFW, Einasto) to describe the cluster's inner matter distribution.

The so-called 2-halo term accounts for lensing from large-scale structure correlated with the cluster, i.e., neighboring halos and surrounding matter that contribute to the signal on larger scales. This contribution becomes significant and should be accounted for in the modeling of the lensing profile to avoid a biased mass reconstruction,  particularly when considering scales beyond the Mpc.

The 2-halo term is now implemented in \clmm (for both the \ccl and \nc theory backends), based on a model combining halo bias with the linear matter power spectrum, as done in \citet{Oguri2011}, with the excess surface density contribution due to the two-halo term computed as
\begin{equation}
    \Delta\Sigma_{\rm 2h}(R) = \frac{\rho_m(z)b(M)}{(1 + z)^3D_A(z)^2} \int\frac{l\mathrm{d}l}{(2\pi)} P_{\rm mm}(k_l, z)J_2(l\theta).
    \label{eq:2h_term}
\end{equation}
Here $P_{\rm mm}(k_l, z)$ is the linear power spectrum,  $b(M)$ is the halo bias (which should be provided by the user),  $J_2(l\theta)$ is the second-order Bessel function, $D_A(z)$ is the angular diameter distance, and $\rho_m(z)$ is the Universe mean density at redshift $z$. Similarly, the 2-halo surface density, $\Sigma_{\rm 2h}$, is available in the code and computed by replacing $J_2(l\theta)$ by $J_0(l\theta)$ in Equation \eqref{eq:2h_term}.
The contribution from the two-halo term can be visualized in Figure~\ref{fig:2h_miscentering_boost}, in this particular example becoming significant around a few Mpc and dominant over the 1-halo contribution beyond $\sim 10 $~Mpc.

\subsection{Boost factors}
\label{subsec:boost}

Contamination of the source galaxy sample by cluster members or foreground galaxies dilutes the signal, as it includes unlensed galaxies in the radial average of the shear. This attenuation is characterized by the so-called boost factor. In practice, this factor can either be applied as a correction to the measurements by multiplying it to the data vector \citep{Sheldon2004lensing,2008JCAP...08..006M}, or as a correction to the modeling by dividing the theoretical prediction by it \citep{2017MNRAS.469.4899M,McClintock2019mass}.

We expect the boost factor to have a radial dependency as the density of galaxies, and consequently the contamination, increases towards the cluster center. Defining $x = R/R_s$ (with $R_s$ a scale radius), two parametrizations of the boost factor $B(R)$ are implemented in \clmm's \code{utils} module and a dedicated demo notebook is provided to exemplify their usage. The first parameterization follows a power-law model (e.g., \citealp{2017MNRAS.469.4899M})
\begin{equation}
    B(R)= 1 + B_0\; x^\alpha
\end{equation}
where $B_0$, $R_s$ and $\alpha$ are free parameters of the model (with $\alpha < 0$ for the boost to decrease with radius and tend to 1 at large distances from the cluster center). The second option follows an NFW excess surface density parametrization \citep{McClintock2019mass}, with only two free parameters $B_0$ and $R_s$, such as
\begin{equation}
    B(x) = 1 + B_0\frac{1-F(x)}{x^2 -1}
\end{equation}
where
\begin{equation}
    F(x) = \begin{cases}
    \frac{\arctan \sqrt{x^2 -1 }}{\sqrt{x^2 -1 }} & \text{if $x>1$}, \\
    \text{1} & \text{if x = 1}, \\
    \frac{\text{arctanh} \sqrt{1-x^2 }}{\sqrt{1- x^2 }} & \text{if $x<1$}.
\end{cases}
\end{equation}
The free parameters in these models are generally not known a priori and can be treated as nuisance parameters, to be fitted alongside the parameters of interest in the inference procedure.
Figure \ref{fig:2h_miscentering_boost} illustrates the dilution of the theoretical signal using an NFW boost model (dotted orange line).

\begin{figure}
    \centering
    \includegraphics{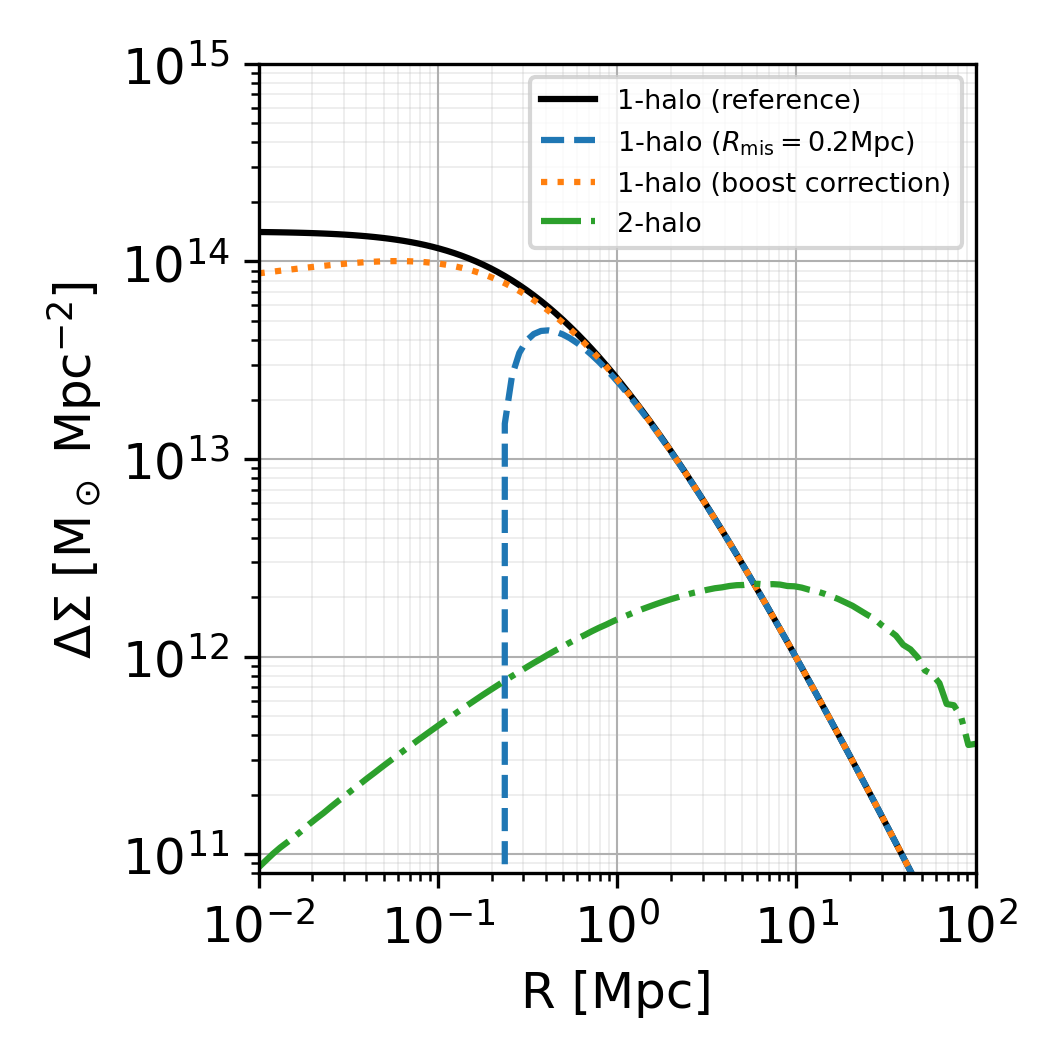}
    \caption{Illustration of the new profile theory functionalities implemented in this version of the code. The solid black line corresponds to the $\Delta\Sigma$ profile of the reference halo defined in Section \ref{sec:intro} ($M_{200,m} = 10^{14}$~M$_\odot$ at $z=0.4$). Assuming a miscentering distance $R_{\rm mis}=0.2$~Mpc moves the profile prediction to the dashed blue line. The model can also be corrected by a boost factor to account for the dilution of the signal by cluster galaxies (the dashed orange line uses an NFW boost model). The 2-halo term, using a characteristic halo bias value $b(M)=2.4$ (dashed-dotted green line), starts dominating the signal after a few Mpc.
    }
    \label{fig:2h_miscentering_boost}
\end{figure}

\subsection{Miscentering}
\label{subsec:miscenter}

The tangential shear profile of a galaxy cluster is measured around a selected density center. For optically-selected clusters, a common choice is to use the position of the  BCG as the cluster center, but this selection may not always be accurate. Identification of the cluster center may differ from one cluster-finding approach to another, even more so when considering different wavelengths (e.g., \citealp{2019MNRAS.487.2578Z,2023A&A...671A..57S,2023arXiv231013207K}). Miscentering will decrease the measured lensing signal in the inner regions and, if not accounted for in the modeling, yield a mass estimate biased low.

Assuming a spherical halo with a projected surface density $\Sigma$ measured from the true center, then the surface density $\Sigma_{\rm mis}(R, R_{\rm mis})$ measured from a miscenter offset distance $R_{\rm mis}$ reads
\begin{equation}
\Sigma_{\rm mis}(R, R_{\rm mis}) = \frac{1}{2\pi}\int_0^{2\pi} \dd\theta \, \Sigma(R')\;,
\end{equation}
where $R'= \sqrt{R^2 + R_{\rm mis}^2 - 2 R R_{\rm mis}}\cos\theta$.
The excess surface density $\Delta\Sigma$ is then computed the usual way as
\begin{equation}
    \Delta\Sigma_{\rm mis}(R, R_{\rm mis}) = {\bar\Sigma_{\rm mis}}(<R, R_{\rm mis}) - \Sigma_{\rm mis}(R, R_{\rm mis})\;,
\label{eq:miscentering}
\end{equation}
with $\bar\Sigma_{\rm mis}(<R, R_{\rm mis})$ the mean surface density inside the projected radius $R$.

The expression above has been implemented in the theory module of \clmm, and the (excess) surface profile functions now allow for a \code{rmis} optional keyword that introduces the computation for a given miscentering distance. Figure~\ref{fig:2h_miscentering_boost} highlights the impact of a 0.2\Mpc miscentering on the $\Delta\Sigma$ profile of a $10^{14}$ M$_\odot$ halo located at $z=0.4$ (solid black versus dashed blue lines).

For a stack weak lensing profile of a cluster ensemble, Equation~\eqref{eq:miscentering} should be further integrated over a miscentering distribution (e.g., \citealp{Oguri2011,Murata2018lensing,2023arXiv230913025S}). For this version of \clmm, only the single halo computation above is provided. This choice is made in the context of the broader DESC cluster cosmology pipeline, currently under development, where we anticipate the integration over the miscentering distribution to happen in another part of the pipeline.

\subsection{Magnification bias}
\label{subsec:mag_bias}
Shear is the most commonly used observable in galaxy cluster lensing, but magnification provides an alternative approach which, despite a lower signal-to-noise, is affected by different systematics and is therefore a complementary valuable probe (e.g., \citealp{2011ApJ...729..127U}).

The magnification of the sources causes a shift on the observed number of sources $n$ below a magnitude limit $m$ in comparison with the true number of sources $n_0$. This effect can be characterized by a magnification bias ($\delta_{\mu}$), defined by $n\left(<m\right)=\left(1+\delta_{\mu}\right)n_0\left(<m\right)$. In \clmm v2.0, we include a parametrization of this bias as a function of the magnification $\mu$ (whose prediction was already included in v1.0):

\begin{equation}
    \delta_{\mu} = \mu^{\alpha - 1}.
\end{equation}
Here, $\alpha$ is the slope parameter linked to the source luminosity distribution as :
\begin{equation}
   \alpha(m) = 2.5 \frac{d}{dm} \log n_0 (<m).
\end{equation}
As we can see in Figure \ref{fig:mag_bias}, the magnification bias can be either positive or negative, depending on the $\alpha$ value of the source sample considered.
\begin{figure}
    \centering
    \includegraphics[width=\linewidth]{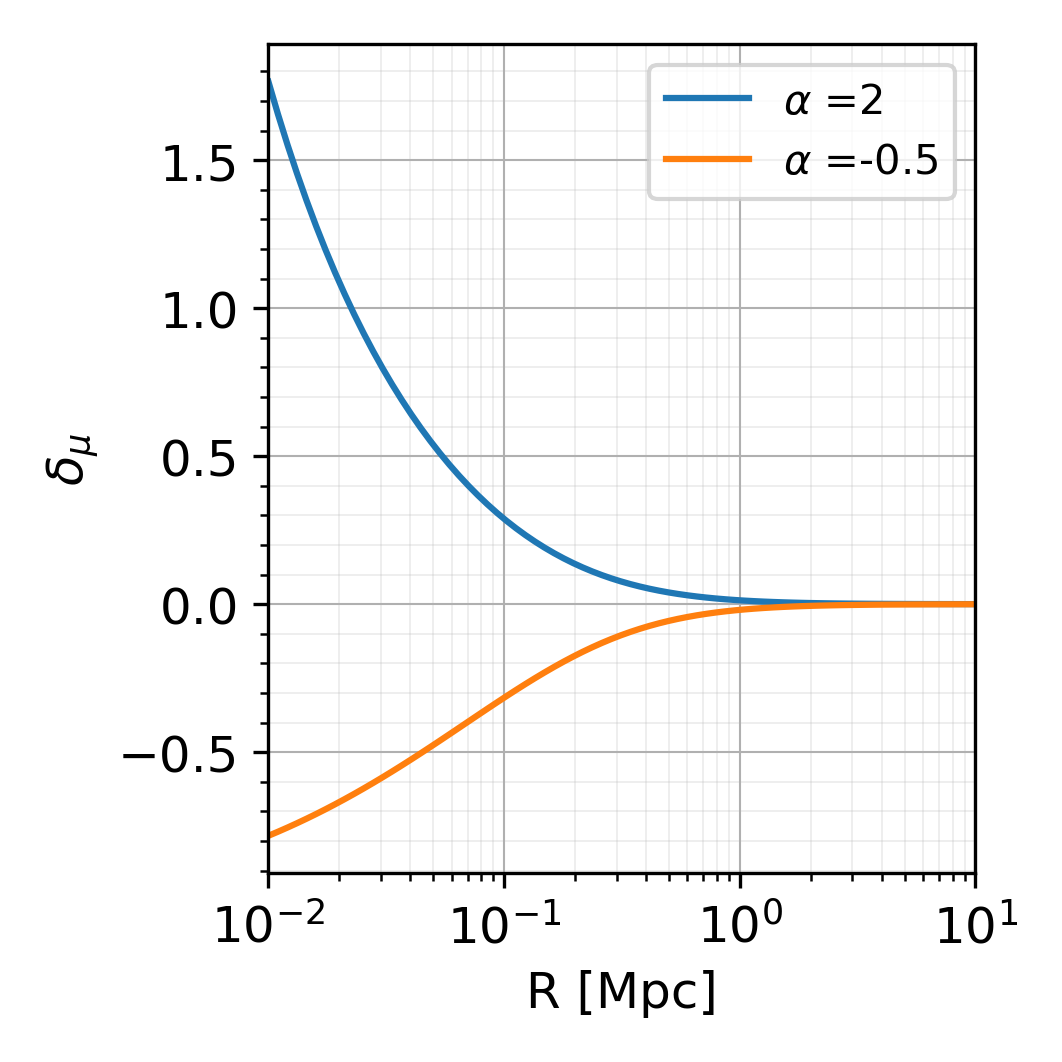}
    \caption{Illustration of the magnification bias profile functionalities implemented in this version of the code. The profiles correspond to a reference halo (see Section \ref{sec:intro}) and a background source sample with different $\alpha$ values, as indicated in the legend.
    }
    \label{fig:mag_bias}
\end{figure}

\subsection{Converting between mass definitions}
\label{subsec:mass_conversion}

The spherical overdensity mass $M_{\Delta}$ is defined according to the spherical overdensity radius $r_{\Delta}$ as :
\begin{align}
    M_{\Delta} = \frac{4\pi}{3}r_{\Delta}^3\rho(z)\;\Delta,
    \label{eq:M_Delta}
\end{align}
with $\rho(z)$ being the mean/critical density of the Universe at the halo's redshift $z$, and $\Delta$ the overdensity scale.
Additionally, the mass within a given radius can be written as a function of the overdensity mass and concentration ($c_\Delta$):
\begin{equation}
    M_{<r}(r, M_{\Delta}, c_{\Delta}) = M_{\Delta}\;\frac{f\left( c_{\Delta} \frac{r}{r_{\Delta}} \right)}{f(c_{\Delta})},
    \label{eq:M_within_r}
\end{equation}
where $r_\Delta$ can be computed from $M_\Delta$ using Equation~\eqref{eq:M_Delta} (and implemented in \clmm as \code{eval\_rdelta}), and $f$ is the volume integral of the halo matter density profile $\rho_{\rm halo}$:

\begin{align}
    f(x)=\int_0^{x} dx'\;x'^2 \rho_{\rm halo}\left( r(x')\right).
    \label{eq:f_Delta}
\end{align}
For some specific profiles, $f$ has an analytical form:
\begin{subnumcases}{f(x)= \label{f_x}}
   \ln(1+x)-\frac{x}{1+x} &\quad \text{NFW}
   \label{eq:3.11a}
   \\
   \gamma(\frac{3}{\alpha}, \frac{2}{\alpha}x^{\alpha})  &\quad \text{Einasto}
   \label{eq:3.11b}
   \\
   \left(\frac{x}{1+x}\right)^2 &\quad \text{Herquist}
   \label{eq:3.11c}
\end{subnumcases}
In Equation~\eqref{eq:3.11b}, $\gamma$ is the lower incomplete gamma function and $\alpha$ is the slope parameter of the Einasto profile.

\newcommand{\DefOne}{_{\Delta_1}^{\text{def}_1}}
\newcommand{\DefTwo}{_{\Delta_2}^{\text{def}_2}}

The conversion between different spherical overdensity definitions relies on the fact that the two definitions should predict the same enclosed mass within the same radius. Therefore, conversion from ($M\DefOne$,  $c\DefOne$) to ($M\DefTwo$,  $c\DefTwo$) can be done by solving the system of equations
\begin{align}
    \left\{
        \begin{array}{lcr}
   M\DefOne = M_{<r}^{\text{def}_1}(r\DefOne, M\DefTwo, c\DefTwo)
   \\
   M\DefTwo =
   M_{<r}^{\text{def}_2}(r\DefTwo, M\DefOne, c\DefOne),
        \end{array}
    \right.
   \label{eq:mass_conversion}
\end{align}
where def$_2$ related terms can all be obtained from $M\DefTwo$ and $c\DefTwo$ using Equation~\eqref{eq:M_within_r} for  $M_{<r}^{\text{def}}$ and Equation~\eqref{eq:M_Delta} to compute $r_{\Delta}^{\text{def}}$ from $M_{\Delta}^{\text{def}}$.

All three profiles of Equations~\eqref{eq:3.11a}-\eqref{eq:3.11c} are supported by the CCL and NumCosmo backends; however, Cluster Toolkit only supports the NFW profile.
Figure~\ref{fig:mass_conversion} shows a comparison between the different mass/concentration definitions.
    In this figure, we assume a reference halo (indicated by the red lines) at $z=0.4$ that has $c_{200,m}=4$, $M_{200,m}=10^{14}M_\odot$ and whose matter distribution follows an NFW profile. Then, we compute the corresponding mass/concentration of that same halo when changing each definition independently and display them with different markers: filled/empty markers indicate $\rho$ definition changes, colors indicate $\Delta$ changes, and markers indicate profile definition changes.

    When we fix $\rho=\rho_{m}, \Delta=200$ (filled orange markers), the mass/concentration values do not change with profile definition. When we only change the profile definition (fix color and filling), the masses remain the same. This happens because the masses are the anchors of the conversions. We also note that, in those cases, the concentrations from NFW are always close to Einasto and higher than Hernquist. Additionally, we can verify that $M_{150}>M_{200}>M_{500}$ (blue>orange>green) and $M_{\Delta,m}>M_{\Delta,c}$ (mass of filled markers>mass of empty markers).

\begin{figure}
 \label{fig:mass_conversion}
    \includegraphics{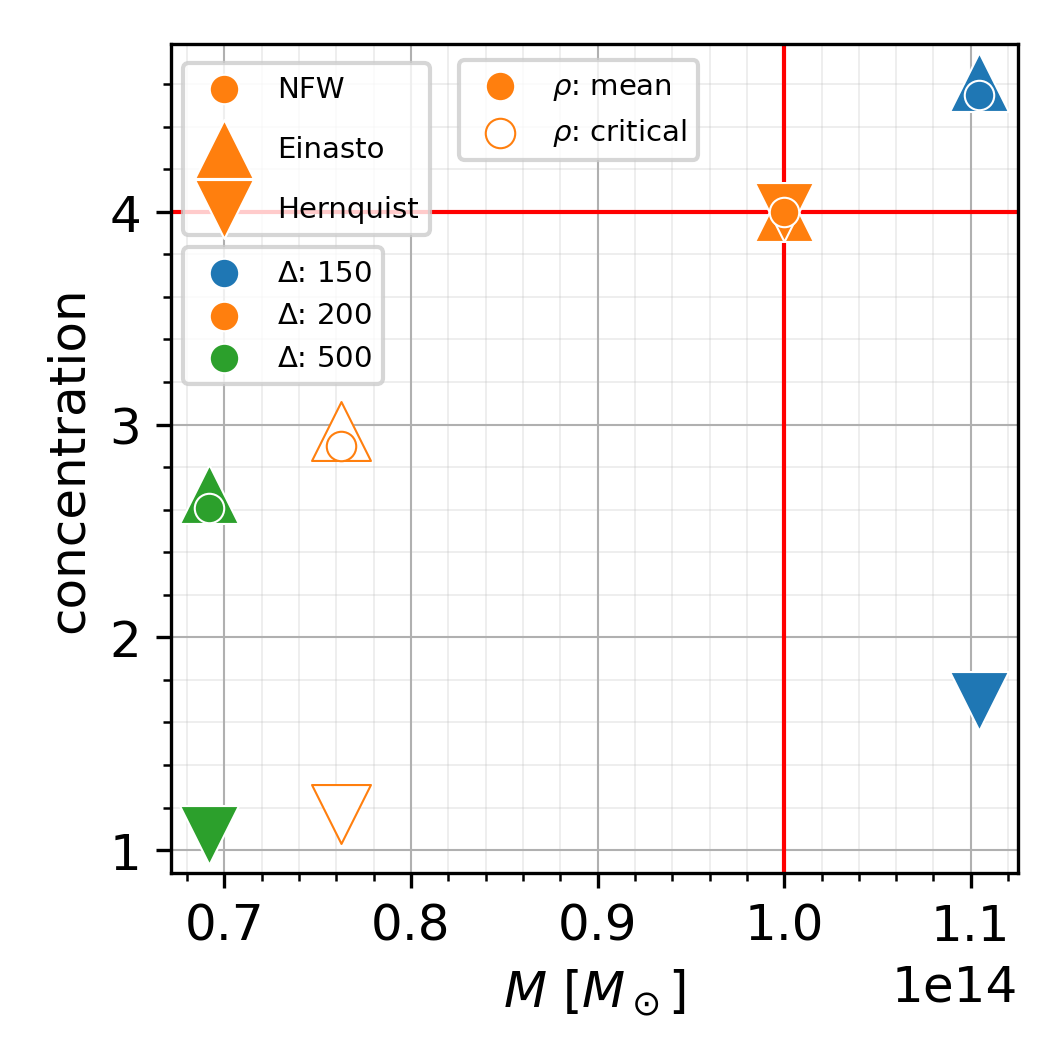}
    \caption{
    Illustration of the mass and concentration conversion: values for our reference halo (indicated by the red lines), given different definitions.
The different symbols indicate a change in the profile definition (NFW, Einasto, or Hernquist), the colors indicate different overdensity contrast values $\Delta$, and the open or filled symbols show values expressed as a function of the critical or the mean density of the Universe.
This figure reads as follows: the mass and concentration of our reference halo can be converted to $M_{200,c}=0.76\times10^{14}M_\odot$ and $c_{200,c}=1.2$ when shifting to a Hernquist profile parametrisation, or to $M_{500,m}=0.69\times10^{14}M_\odot$ and $c_{500,m}=2.6$ when shifting to an Einasto profile. }

\end{figure}

\subsection{Handling different redshift information}
\label{subsec:redshift}

\
In order to account for different use cases, we allow for different ways for the user to provide information about the redshift of sources in our modeling of the convergence and tangential shear, as well as reduced tangential shear, magnification, and magnification profile. The redshift information can thus be given either as:
\begin{itemize}
    \item \emph{discrete values}, $\{z_1,..., z_n\}$ or $z$=const.: in the case of individual source redshifts or all sources in a single plane;
    \item \emph{a distribution function}, $n(z)$: in the case where the global source redshift distribution is known;
    \item \emph{averaged geometrical lensing efficiency}, $\mean{\beta_s}$ \& $\mean{\beta_s^2}$ (essential to model the reduced tangential shear, see next section): in the case these parameters are known.
        These quantities are computed from (i) the source redshift-dependent geometrical lensing efficiency $\beta_s(z)=\Sigma_{\rm crit}(z\rightarrow \infty)/\Sigma_{\rm crit}(z)$,
        where $\Sigma_\mathrm{crit}$ is the critical surface mass density (see Equation~{19} in \paperI).
        and (ii) the source redshift distribution $n(z)$ such as
        \begin{equation}
             \mean{\beta_s^\alpha} = \frac{\int n(z)\beta_s^\alpha(z)\, dz}{\int n(z)\, dz}.
        \end{equation}
    In this approach, the reduced tangential shear and the magnification and magnification bias can only be estimated with approximated relations. The first- and second-order approximations are shown in Section \ref{subsec:approx}. This case can be very useful, as it produces a good approximation of the full modeling of $n(z)$ with a much lower computational processing time, and allows comparing more easily to studies for which only those parameters are reported.
\end{itemize}

Figure \ref{fig:diff_z_type} illustrates an example of the equivalent input for each type of redshift information.

\begin{figure}
    \centering
    \includegraphics[width=\linewidth]{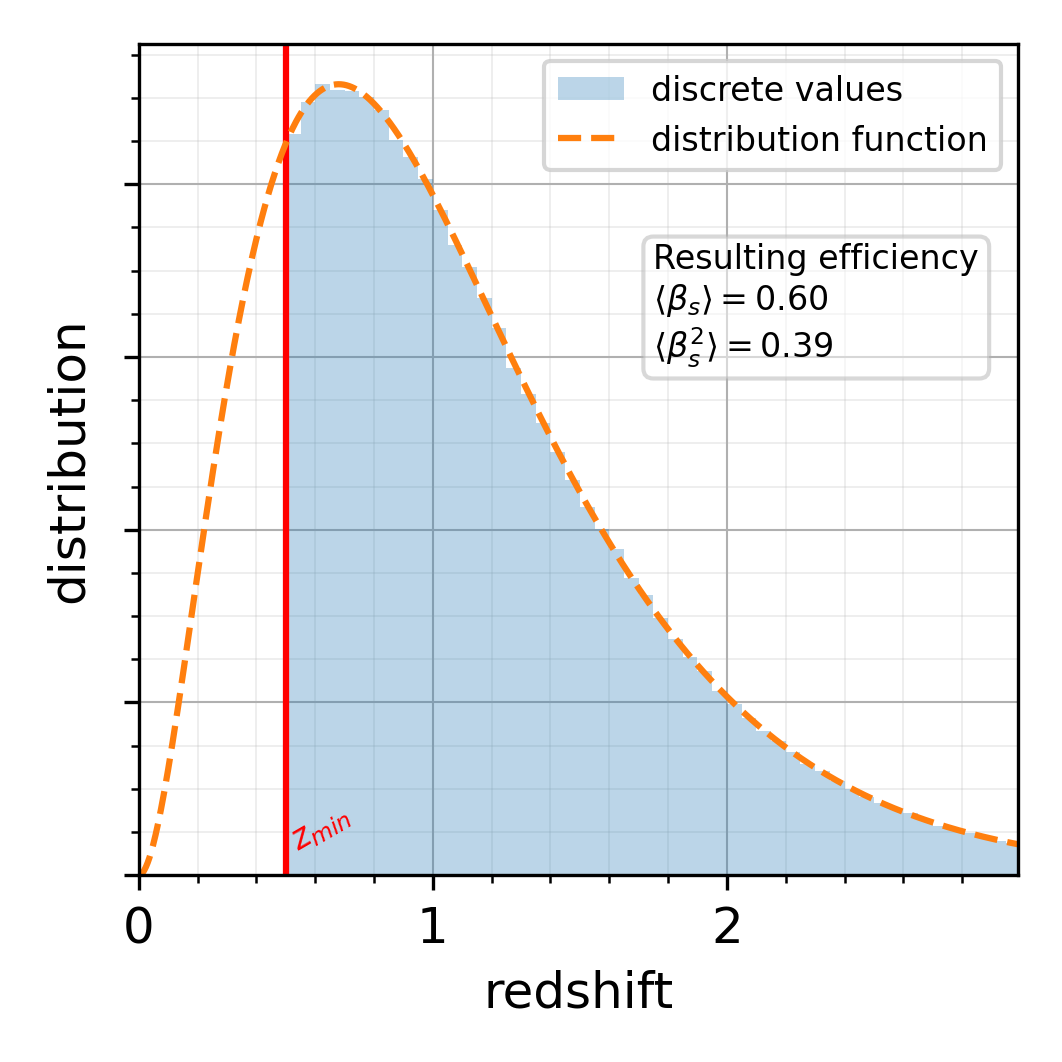}
    \caption{Illustration of the different source redshift information that can be used to predict cluster lensing quantity profiles. The orange dashed line shows a given form of the source redshift distribution, the blue shaded region shows the histogram of individual sources' redshifts that follow that distribution, and the corresponding averaged lensing efficiency parameters are indicated in the additional box. The minimum redshift (corresponding to the cluster redshift plus a given step) is indicated by the red line.
    }
    \label{fig:diff_z_type}
\end{figure}

\subsection{Approximations for profiles}
\label{subsec:approx}

When the averaged lensing efficiency parameters are used as the redshift information (see Section \ref{subsec:redshift}), the reduced tangential shear, magnification, and magnification bias can only be estimated with approximated relations. Here we present the first- and second-order approximations of these quantities.

The geometrical lensing efficiency $\beta_s$ can be used to compute the reduced shear and convergence as a function of their quantities at $z\rightarrow \infty$: $\gamma(z)\rightarrow\beta_s(z)\gamma_\infty$, $\kappa(z)\rightarrow\beta_s(z)\kappa_\infty$. Using this transformation, the reduced tangential shear, the magnification, and the magnification of each individual source can be described as a function of their $\beta_s$:

\begin{equation}
    g(z_s)
    =\frac{\beta_s(z)\gamma_{\infty}}{1-\beta_s(z)\kappa_{\infty}},
	\label{eq:g_individual}
\end{equation}
\begin{equation}
\mu(z)
= \frac{1}{\left[1-\beta_s(z)\kappa_{\infty}\right]^2-\left[\beta_s(z)\gamma_{\infty}\right]^2}
\label{eq:kappa}
\end{equation}
and
\begin{equation}
\delta_\mu(z)=
\mu^{\alpha-1}(z)
\label{eq:kappa_bias}
\end{equation}
respectively.

The theoretical prediction of those quantities for a given lens can be computed based on their mean value, given the sources redshift. This computation depends on the type of input source redshift information.
In the case of discrete sources, the mean value of these quantities is simply computed as the arithmetic mean of individual values: $\mean{q}=\sum q_i/N$.
For sources with a known redshift distribution function $n(z)$, the mean value of a quantity $q$ can be computed as $\mean{q}=\int q(z) n(z)\mathrm{d}z/\int n(z)\mathrm{d}z$.

When the averaged lensing efficiency $\mean{\beta_s}$ and $\mean{\beta_s^2}$ are given as the redshift information, the mean values of ($g, \mu, \delta_\mu$) can be estimated based on approximations of Equations~\eqref{eq:g_individual}-\eqref{eq:kappa_bias}. This can be done with the \texttt{approximation} argument in \clmm.

\newcommand{\orderI}[1]{#1^{\texttt{type1}}}
\newcommand{\orderII}[1]{#1^{\texttt{type2}}}

For the reduced shear, two approximations are made available, which correspond to \cite{applegate14} and \cite{schrabback_2017} respectively:

\begin{equation}
    \mean{{\orderI{g}}}=
    \frac{\mean{\beta_s}\gamma_{\infty}}{1-\frac{\mean{\beta_s}^2}{\mean{\beta_s}}\kappa_{\infty}},
	\label{eq:g_order1}
\end{equation}

\begin{equation}
\mean{\orderII{g}}=
    \frac{\mean{\beta_s}\gamma_{\infty}}{1-\mean{\beta_s}\kappa_{\infty}} \left[1+\left(\frac{\mean{\beta_s^2}}{\mean{\beta_s}^2}-1\right)\mean{\beta_s}\kappa_{\infty}\right].
\label{eq:g_order2}
\end{equation}

\noindent
This option was previously implemented as the \code{Applegate14} and \code{Schrabback18} approximation in the theory module.

For the magnification and magnification bias, the \code{approximation} option can be interpreted as the order of the approximation, with the first order being:

\begin{equation}
\begin{split}
\mean{\orderI{\mu}}
&\approx1+2\mean{\beta_s}\kappa_{\infty},
\end{split}
\label{eq:kappa_order1}
\end{equation}

\begin{equation}
\mean{\orderI{\delta_\mu}}
\approx1+2(\alpha-1)\mean{\beta_s}\kappa_{\infty}.
\label{eq:kappa_bias_order1}
\end{equation}

\noindent
and the estimation of the second-order approximation is:

\begin{equation}
\begin{split}
\mean{\orderII{\mu}}
&\approx1+2\mean{\beta_s}\kappa_{\infty}
+3\mean{\beta_s^2}\kappa_{\infty}^2+\mean{\beta_s^2}\gamma_{\infty}^2,
\label{eq:kappa_order2}
\end{split}
\end{equation}

\begin{equation}
\begin{split}
\mean{\orderII{\delta_\mu}}
\approx1&+2(\alpha-1)\mean{\beta_s}\kappa_{\infty}\\
&+(\alpha-1)\mean{\beta_s^2}\gamma_{\infty}^2\\
&+(2\alpha-1)(\alpha-1)\mean{\beta_s^2}\kappa_{\infty}^2,
\label{eq:kappa_bias_order2}
\end{split}
\end{equation}
respectively.

These approximated calculations involving averaged lensing efficiency in this subsection can also be used with a redshift distribution function input. In this case, the values $\mean{\beta_s}$ and $\mean{\beta_s^2}$ are computed with the integral of the distribution function.

An example of reduced shear and magnification profiles for the different approximation cases are shown in Figure \ref{fig:approx_profiles} for our reference halo. Here the difference between the approximated profiles is below 1\% in the radius range shown. Note however that this difference increases strongly with the mass of the halo considered.

\begin{figure*}
    \includegraphics{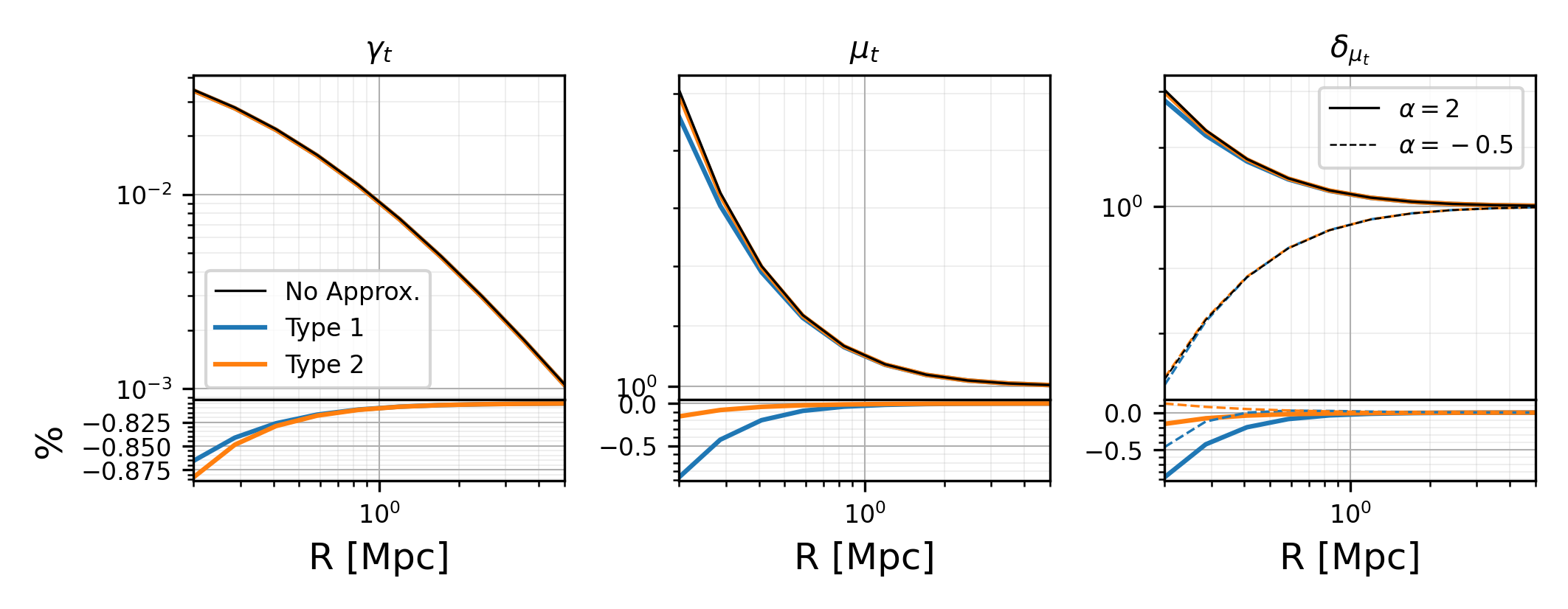}
    \caption{Plots illustrating the tangential shear profile, magnification, and magnification bias profiles of our reference halo using the \cite{chang2013effective} $N(z)$ source redshift distribution and corresponding approximations described on Section~\ref{subsec:approx}. Bottom panels show the difference between the exact approach and the approximations. The magnification bias profiles are shown for two different values of the $\alpha$ parameter describing the source sample, as indicated in the legend.
    }
    \label{fig:approx_profiles}
\end{figure*}

\subsection{Halo triaxiality modeling}
\label{subsec:triaxial_mod}
Collapsed dark matter halos are not always expected to be spherical due to their complex individual accretion histories within the cosmic web \citep{Sheth2001hmf,Jing2002dmhalostriaxial,Schneider2012triaxial,Despali2014triaxialsim}. Consequently, assuming clusters to be spherical can introduce significant bias in the estimation of cluster properties from weak-lensing analyses. Various simulation-based studies \citep{Corless2007triaxialhalo,Meneghetti2010Xrays,Becker2011simtriaxiallensing,Lee2018massbiasbaryon,Giocoli2014simtriaxial,Herbonnet2022the300bias} demonstrate that lensing masses inferred under the assumption of sphericity are strongly affected by the triaxial shape of halos, resulting in a mass bias of $\sim$$20\%$  for single clusters. This bias can, however, be reduced to a few percent by stacking the WL signal of clusters \citep[see e.g.][]{McClintock2019mass}.  This remains a major source of systematic uncertainty in interpreting the weak-lensing signal, with inevitable implications for cosmological constraints.

When the projected dark matter distribution deviates from sphericity, the local shear exhibits non-trivial radial and polar dependencies, which can be analyzed through multipole decomposition. The multipolar coefficients of the shear field can be employed to infer the projected ellipticity of galaxy-sized dark matter halos \citep{Clampitt2016lensingellipticity,Robison2023multipole} or the large-scale dark matter distribution surrounding galaxy clusters \citep{Gonzalez2020lensingellipticity,Shin:2017rch}.
The two-dimensional surface mass density of a cluster, $\Sigma(R, \theta)$, can be formally expressed using its multipole expansion \citep{Bernstein2009multipole}:
\begin{align}
    \Sigma(R, \theta) &= \frac{1}{2}\sum_{m=-\infty}^{m=+\infty}(1 + \delta_m)\Sigma^{(m)}(R)e^{im\theta},
    \label{eq:multipole_expansion}\\
    &=\Sigma_0(R) + \Sigma_2 (R) \, \cos (2\theta) + \cdots
\end{align}
where $\delta_m$ is the Kronecker delta.
Here, $\Sigma^{(m)}$ represents the $m$-th multipole coefficient of $\Sigma(R, \theta)$, defined as:
\begin{equation}
    \Sigma^{(m)}(R)=\frac{1}{(1+\delta_m)\pi}\int_0^{2\pi}\Sigma(R, \theta)e^{-im\theta}d\theta.
    \label{eq:multipolem}
\end{equation}
In this formulation, we define $\Sigma_0(R) = \Sigma^{(0)}(R)$ and $\Sigma_2(R) = (1/2)[\Sigma^{(2)}(R)+\Sigma^{(-2)}(R)]$ as the monopole and quadrupole components of the projected surface density, respectively.
    When observing galaxy ellipticities, we rather have access to the excess surface density (or equivalently the shear field $\epsilon_{\rm gal} =\epsilon_{\rm int} +\gamma_1(R,\theta) + i\gamma_2(R,\theta)$, when $|\gamma|\ll 1$) than the surface density (i.e. the convergence field).

    According to \citet{Schneider1991multipole} and \citet{Bernstein2009multipole}, in the weak lensing regime, each multipolar coefficient $\Delta\Sigma^{(m)}$ depends on the multipole $\Sigma^{(m)}$ of the same order $m$  \footnote{Let us note that, in the $not-so-weak$ lensing regime, $\Delta\Sigma^{(m)}$ does depend on all the multipolar coefficients $\Sigma^{(m)}$.}. Following the prescription in  \cite{Shin:2017rch}, we thus define two \textit{optimal} measures of the excess surface density multipole in the weak lensing regime, the first one :

\begin{multline}
\label{eq:triaxiality_4theta}
    \Delta\Sigma^{4\theta} (R) \equiv \dfrac{\Sigma_{\rm crit}}{2\pi} \int \mathrm{d}\theta \left( \gamma_1(R,\theta)\cos{4\theta} + \gamma_2(R,\theta) \sin{4\theta} \right)  \\
    = \dfrac{\Sigma_2}{2} - \dfrac{3}{R^4}\int_0^R \mathrm{d}R' R'^3 \Sigma_2 (R')
\end{multline}
and the second :
\begin{multline}
    \label{eq:triaxiality_const}
    \Delta\Sigma^{\rm const} (R) \equiv \dfrac{\Sigma_{\rm crit}}{2\pi} \int \mathrm{d}\theta \gamma_1(R,\theta) \\
    = \dfrac{\Sigma_2}{2} - \int_R^{\infty} \mathrm{d} R' \dfrac{\Sigma_2(R')}{R'}.
\end{multline}
We refer the readers to \cite{Adhikari2015} and \cite{Shin:2017rch} for details of this derivation.

Below, we will consider the case of an elliptical lens without miscentering
\footnote{Let us note that even for a spherical cluster mass distribution the multipole decomposition above is still valid, when having an offset between the \textit{true} cluster center and the observed one (as detailed in Section~\ref{subsec:miscenter}). In this case, the 2D surface density is a non-trivial function of the projected distance $R$ from the wrong center and the polar angle $\phi$, and a lensing dipole needs to be considered.}.
Following \citet{Adhikari2015} with a slightly changed formalism, the surface density is well described by two-dimensional elliptical iso-density contours, i.e., $\Sigma(R,\theta)\propto \tilde{R}^{\tilde{\eta}}$, where $\tilde{R} = (q x^2 + y^2/q)^{1/2}$}, where $q$ is the axis ratio. For small deformations (i.e., $q$ close to 1), we get \citep[see also][]{2026arXiv260506587F}:

\begin{multline}
    \Sigma_0(R) = \Sigma_{\rm sph}(R)\left[1 + \dfrac{\epsilon^2}{2} \left(\eta(R) + \dfrac{\eta(R)^2}{2} + \frac{1}{2}\dfrac{\mathrm{d} \eta(R)}{\mathrm{d}\ln R} \right) \right] \\
    + \mathcal{O}( \epsilon^4 )
\label{eq:triaxiality_monopole}
\end{multline}
where $\Sigma_{\rm sph}(R)$ is the spherical projected surface density with choice of NFW, Einasto, or Hernquist profiles, and $\eta$ is defined as the logarithmic slope of $\Sigma_{\rm sph}(R)$,
\begin{equation}
    \eta(R) = \frac{\mathrm{d} \ln{\Sigma_{\rm sph}(R)}} {\mathrm{d} \ln{R}},
\end{equation}
\noindent
and we have defined the 2D ellipticity of the halo as:

\begin{equation}
    \epsilon = \dfrac{1-q}{1+q}.
\end{equation}
Moreover, we find that :
\begin{equation}
    \Sigma_2(R) = - \epsilon \eta(R) \Sigma_{\rm sph} (R) \cos({2\theta_0}) + \mathcal{O}( \epsilon^3 ),
\end{equation}
where $\theta_0$ is the polar angle of the major axis ($\theta_0=0$ when we align the major axis to the $x$-axis).
Note that, assuming an elliptical cluster, Equations~\eqref{eq:triaxiality_4theta} and~\eqref{eq:triaxiality_const} asymmetrically depend on $\gamma_1$ and $\gamma_2$ in the Cartesian coordinates, meaning that a specific orientation of the halo is assumed.
Specifically, it is assumed here that the semi-major axis is aligned with the $x$-axis.
Because of this, a rotation by some angle $\phi$ must be applied to the lensing signal and the Cartesian coordinates by the user.
This may, for instance, be taken as either the BCG orientation or the orientation determined from the distribution of lens member galaxies.
This is critically important to consider when stacking galaxy clusters (described in Sect.\ref{sec:new_data}) since stacking misaligned halos will cause a rounding-out of the ellipticity. In practice, one may also need to select clusters with high centering quality to reduce miscentering effects.

\begin{figure*}[htbp]
    \centering
    \includegraphics{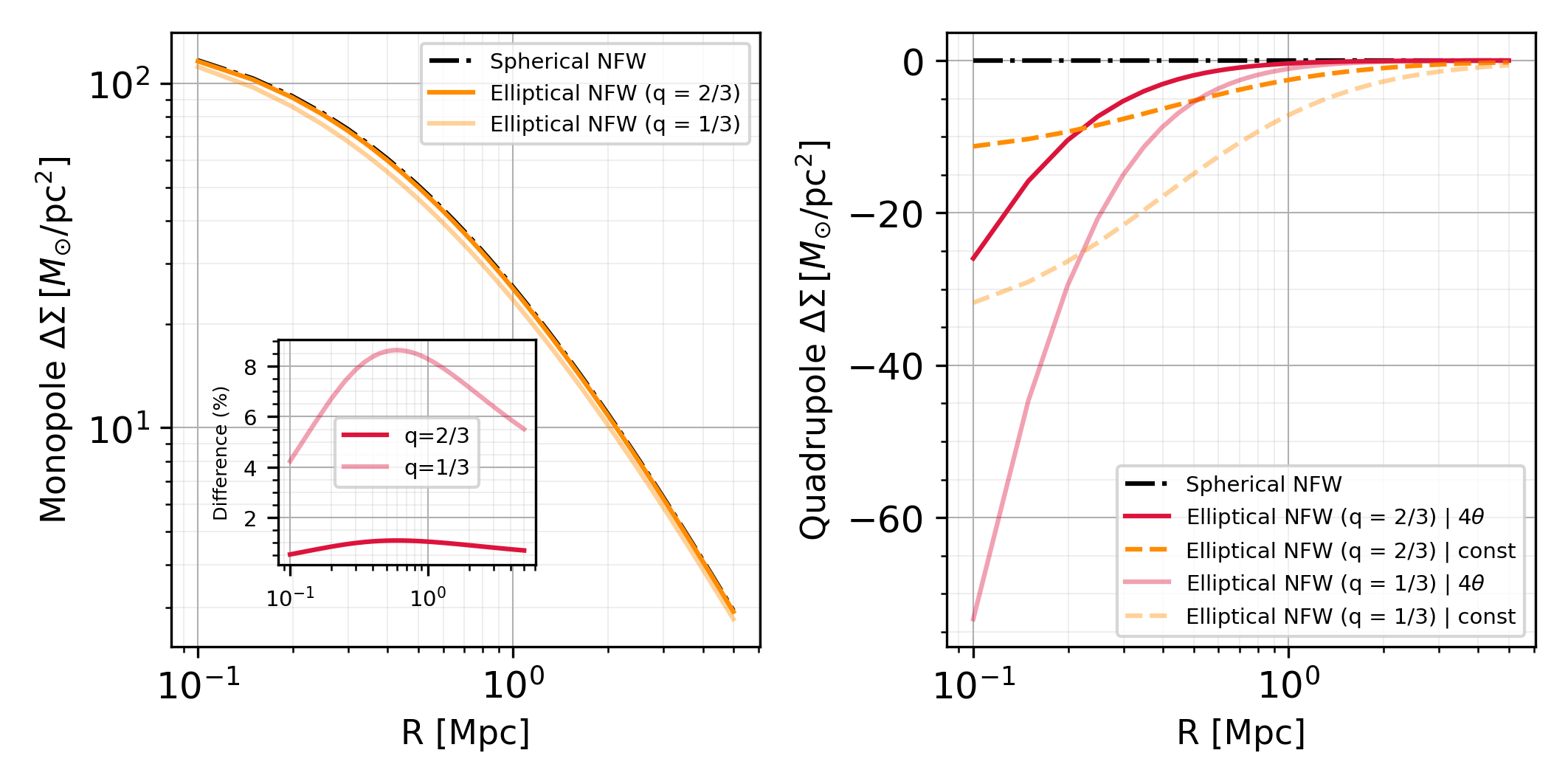}
    \caption{
    Lensing signal models for halo triaxiality. We consider the default halo example defined for this work, but with projected axis ratios $q=2/3$ and $q=1/3$ ~\citep{2024MNRAS.530.2251F}. \textit{Left}: spherical NFW profile and monopole signals (the $\lesssim1\%$  difference at $q=2/3$ comes from the $e^2$ term in Equation~\eqref{eq:triaxiality_monopole}). The inset shows the percentage difference between the spherical NFW and the monopole profiles.
    \textit{Right}: quadrupole signals for the two estimators (Equations~\eqref{eq:triaxiality_4theta} and~\eqref{eq:triaxiality_const}). We use the lighter shading for the $q=1/3$ halo.
    }
    \label{fig:triaxiality}
\end{figure*}

In this new version of \clmm, we implemented the leading-order corrections to the surface density to account for halo triaxiality, and we show in Figure~\ref{fig:triaxiality} the monopole and quadrupole models. Note that the amplitudes of both quadrupoles grow with ellipticity and concentration.

\section{Update and new functionalities for cluster weak lensing measurements}
\label{sec:new_data}
In this section, we present new functionalities and updates on the operations related to data measurements. This includes updates on the source galaxy catalog description and handling, new estimators of the lensing signal, and a new class for cluster stacking purposes.

\subsection{Coordinate Systems of the source catalog}
\label{subsec:coord}
Shear measurements can be defined in different coordinate systems. As we aim to be able to ingest different input types, we now allow the user to specify the coordinate system of the shear data. The two supported options are \code{celestial} and \code{euclidean}. Following \cite{rowe2015galsim}, we define both as follows:

\begin{itemize}
    \item Celestial coordinates are defined with the declination $\delta$ taking the role of the polar angle, and the right ascension $\alpha$ functioning as the azimuthal angle.
    \item Euclidean coordinates are defined as a Cartesian system in a plane tangent to the celestial sphere at the point of observation. At this point of contact, the y-axis is taken as parallel to the declination and the x-axis as antiparallel to the right ascension.
\end{itemize}

In small angles, we can use a planar approximation of the celestial coordinates, and both systems are related by a parity transformation of the x-axis. Even though the shear measurements in both systems are different, the conversion is trivial. Given a polar angle $\phi$ in the plane of the image, the transformation from $\epsilon^E_1 + i \epsilon_2^E = |\epsilon| e^{2i\phi}$ to $\epsilon^C_1 + i \epsilon_2^C = |\epsilon|e^{2i\phi^\prime}$ is given simply by $\phi \rightarrow \pi - \phi$:

\begin{equation}
    \epsilon_1^C + i\epsilon_2^C = \epsilon_1^E - i\epsilon_2^E.
\end{equation}

In the previous version of \clmm, we expected the user to provide shear data in Euclidean coordinates.
Now, the user is allowed to set the defined coordinate systems at three points: on the creation of a \code{GCData} object from a provided catalog; on the generation of a mock catalog; when calling standalone functions. In the first two options, coordinate system information is stored in the \code{GCData} metadata and can be updated and converted at any time. This metadata is also used by the \code{GalaxyCluster} object to perform all necessary calculations.

Finally, we also supply a notebook
\footnote{\url{https://github.com/LSSTDESC/CLMM/blob/main/examples/demo_coordinate_system.ipynb}}
 with examples of correctly and incorrectly supplied coordinate systems, coordinate system conversion, and mock data generation.

\subsection{Source redshift inputs}
\label{subsec:z_input}
\

In the new version of \clmm, the \code{GCData} object has been set up to accept
a larger variety of types for the photometric redshift information of the lensed sources.
These are determined by setting the type of redshift input to the function \code{CGData.pzpdf\_info["type"]}, along with the different redshift input. The current options are:
\begin{itemize}
    \item Point source estimations:
        \code{type} must be \code{None} and the redshift value must be provided in column \code{z};
    \item Posterior distribution function (PDF) with individual binning for each source:
        \code{type} must be \code{"individual\_bins"}, the individual range of each PDF must be provided to the column \code{pzbins} and the PDFs in column \code{pzpdf};
    \item PDF with same binning for all sources:
        \code{type} must be \code{"shared\_bins"}, the  unique array with a redshift range must be added to \code{CGData.pzpdf\_info["zbins"]} and the PDFs in column \code{pzpdf};
    \item quantiles of the PDF:
        \code{type} must be \code{"quantiles"}, the set of quantiles (ex: \code{[0.16, 0.50, 0.84]}) must be added to \code{CGData.pzpdf\_info["zbins"]}, and the corresponding redshift locations must be added to the column \code{pzquantiles}.
\end{itemize}

\noindent
On the cases where the PDF information of the sources is provided, \clmm\ can output the PDF on a specified redshift grid, which is used to compute the background probability and the weights for stacking clusters. This conversion is based on the LSST-DESC package \qp\footnote{\url{https://github.com/LSSTDESC/qp}}.

\subsection{Background probability for source selection}
\label{subsec:backprob}

Sharp cuts based on magnitudes and/or redshift can be used to ensure the source sample used is pure. However, when the sources contain photometric redshifts, we can compute the probability of a given galaxy being a background galaxy by weighting the contribution of the PDF for each source $p_{s}(z)$ behind the cluster:
\begin{equation}
    P_{ls}^{\rm bkg}= \int_{z_l}^{\infty} p_s(z) dz.
    \label{eq:prob_bkg}
\end{equation}

\noindent We implemented this functionality in \clmm, so that the user can set a background probability threshold to ensure the selection of sources results in a clean selection. This quantity can be computed with the \code{compute\_background\_probability} function available in the \code{dataops} module and in the \code{GalaxyCluster} object.

\subsection{Lens-source weights}
\label{subsec:weights}

\newcommand{\ShapeNoise}{\sigma_{\rm shape}}
\newcommand{\ShapeError}{\sigma_{\rm meas}}
\newcommand{\wsig}{w_{ls}^{(\Delta\Sigma)}}

The effective contribution of each source to the weak lensing signal around a cluster (and lens-source pair contributions for cluster stacks) can be accounted for with the use of weights. These lens-source weights can be computed as the square inverse of the source shape uncertainty \citep{Sheldon2004lensing},
which is composed of the intrinsic dispersion of galaxy shapes ($\ShapeNoise$) and the shape measurement errors ($\ShapeError$):
\begin{equation}
    \label{eq:weight_shape}
    w_{ls} = \frac{1}{\ShapeNoise^2 + \ShapeError^2}.
\end{equation}

In the case of the excess surface density $\Delta\Sigma$, the weight also has an additional geometric contribution from the effective surface density $\SigCritEff$ \citep{Sheldon2004lensing}:

\begin{equation}
    \label{eq:weights}
    \wsig = w_{ls}^{\rm geo}\times w_{ls}
    =\frac{w_{ls}}{(\SigCritEff)^2},
\end{equation}

\noindent
which is defined by:

\begin{equation}
    \SigCritEff = \left[\int_{z_l}^{+\infty}\Sigma_{\rm crit}(z_l, z)^{-1}p_s(z) dz\right]^{-1},
    \label{eq:sig_eff}
\end{equation}

\noindent
where $p_{s}(z)$ is the source redshift PDF. When using point estimates for the source redshifts, the geometrical weight is equal to

\begin{equation}
    w_{ls}^{\rm geo}= \Sigma_{\rm crit}^{-2}(z_l, z_s).
    \label{eq:w_geo}
\end{equation}

These different weights can be computed with the \code{compute\_galaxy\_weights} function available in the \code{dataops} module and in the \code{GalaxyCluster} object. Depending on the value of the flag \code{is\_deltasigma}, Eq. \ref{eq:weight_shape} or Eq. \ref{eq:weights} will be returned.

\subsection{Halo triaxiality measurements}
\label{subsec:triaxial_meas}

The two optimal measures for triaxiality of the excess surface
density multipoles (the data counterparts of Equations~(\ref{eq:triaxiality_4theta}, \ref{eq:triaxiality_const}) respectively) for a given lens $l$, are computed using \citet{Shin:2017rch}:

\begin{equation}
    \Delta\Sigma^{4\theta}_l  = \frac{\sum_s \wsig\left( \epsilon^{\rm obs}_{1,s}\cos{4\theta_{s}} +  \epsilon^{\rm obs}_{2,s}\sin{4\theta_{s}} \right) \SigCritEff}{\sum_s \wsig} ,
    \label{eq:delta_sigma_4theta_est}
\end{equation}

\noindent and
\begin{equation}
    \Delta\Sigma^{\rm const}_l =  \frac{\sum_s \wsig \epsilon^{\rm obs}_{1,s} \SigCritEff}{\sum_s \wsig},
\end{equation}

\noindent
where
$\theta_s$ is the angular separation between the source galaxy and the reference center of the cluster,
$\epsilon^{\rm obs}_{s}=\epsilon^{\rm obs}_{1,s}+i\epsilon^{\rm obs}_{2,s}$ is galaxy ellipticity,
$\wsig$ and $\SigCritEff$ are obtained by Equations~\ref{eq:weights}-\ref{eq:sig_eff} respectively, and the sums run over the sources $s$.

These terms are optionally computed by the \code{compute\_tangential\_and\_cross\_components} function available in the \code{dataops} module and in the \code{GalaxyCluster} object.

\subsection{The {\rm\code{ClusterEnsemble}} Object}
\label{subsec:clusterensemble}

While it is possible to obtain precise mass measurements for massive clusters up to intermediate redshifts, which contain a large number of sources with reasonable shear signals, most of the optical clusters detected on photometric surveys will be restricted to noise-dominated sources. Stacking galaxy clusters in redshift and mass-proxy (e.g., richness) bins allows us to compensate for this signal-to-noise limitation in mass measurements.
The stacked weak lensing quantity profile $\Theta$ (where $\Theta$ can stand for $\gamma_t$, $\kappa$, $\Delta\Sigma$, ...) is computed as :
\begin{equation}
    \widehat{\Theta}(R) = \frac{\sum\limits_{l} w_{l}(R) \Theta_{l}(R)}{\sum\limits_{l} w_{l}(R)},
    \label{eq:theta_hat_stacking}
\end{equation}
with the index $l$ standing for each individual lens, and $w_{l}(R)=\sum_s w_{ls}(R)$ is the sum of the weights $w_{s,l}$ over all sources $s$ between $R$ and $R+\delta R$ (see Section \ref{subsec:weights}).

The \texttt{ClusterEnsemble} class was created to facilitate the handling of operations on ensembles of galaxy clusters (i.e., clusters in a given redshift and mass proxy bin), which include stacking lensing profiles, estimating their covariance, and saving/loading whole ensembles.
\texttt{ClusterEnsemble} objects hold the main cluster quantities (ex: id, ra, dec, redshift, weights) and the weak lensing stacked profiles (radial coordinates, tangential and cross components, and weights).

\subsection{Covariance estimation}
\label{subsec:cov}
The covariance matrix of a stacked lensing measurement accounts for different contributions \citep{Gruen2015covDS,Wu2019covDeltaSigma}. The two major contributions are the intrinsic scatter in the shape of background galaxies (shape noise) and the limited number of clusters and background galaxies (cluster and galaxy shot noise). The other contributions come from the uncorrelated large-scale structures along the line of sight, which introduce a scatter in the measured convergence. Stochastic variation also arises from correlated halos around the clusters in the sample, associated with the variation of the 2-halo term. Variations of intrinsic halo properties, i.e., the variation of concentration at fixed mass, the halo ellipticity and orientation, also contribute to the total covariance. Finally, it may also denote the scatter in individual masses of richness-selected clusters within the stack.
Additional minor sources of scatter are errors induced by the presence of substructures in halos, the photometric redshift of background galaxies, super-sample covariance, etc.

\newcommand{\pradial}[2][]{\Theta_{#1}(R_{#2})}
\newcommand{\pstack}[2][]{\widehat{\Theta}_{#1}(R_{#2})}
\newcommand{\pstackm}[2][]{\overline{\Theta}_{#1}(R_{#2})}

In practice, all possible contributions to the covariance (even those that are not easy to derive analytically) can be captured using re-sampling techniques on real data.
The measured covariance of a stacked cluster lensing profile between radial bin $i$ and $j$ is given by :
\begin{equation}
    \mathbf{C}^{\rm true}_{ij} = \left\langle
    \left(\pstackm{i}- \pstack{i}\right)
    \left(\pstackm{j}- \pstack{j}\right)
    \right\rangle,
    \label{eq:cov_true}
\end{equation}
where $\langle\cdot\rangle$ denotes the average over the data sample
and $\pstackm{i} = \left\langle\pstack{i} \right\rangle$. In the literature, several methods for the prediction of the covariance have been used, including analytical models \citep{Shirasaki2018cov,Wu2019covDeltaSigma}\footnote{\citep{Wu2019covDeltaSigma} have created the \texttt{cluster-lensing-cov} package. The GitHub repository is available at \url{https://github.com/hywu/cluster-lensing-cov}}, and using accurate mock simulations \citep{Shirasaki2017covarianceDeltaSigma}.

In \clmm, we implemented methods for the estimation of the covariance from the measured profiles, as it is a commonly used techniques in cosmological cluster lensing studies. As mentioned before, this approach captures all possible contributions to the covariance.
The three methods that have been implemented are described in detail below: Jackknife; Bootstrap; Sample Covariance:

\newcommand{\jk}{{\rm JK}}
\newcommand{\bs}{{\rm BS}}
\newcommand{\SC}{{\rm SC}}

(i) The Jackknife covariance matrix is the most widely used technique in cluster lensing cosmology (see \citet{McIntosh2016jacknife} for a pedagogical presentation of the Jackknife method; see e.g. \citet{McIntosh2016jacknife,Escoffier2016jackknife,Shirasaki2017covarianceDeltaSigma,McClintock2019mass} for the application of the Jackknife method in cosmology). It is computed as follows: First, we divide the sky area into $N_\jk$ equal-sized sub-areas. For each index $k\in [1, N_\jk]$ denoting each Jackknife sub-region, we exclude the clusters in the $k$-th region and compute $\widehat{\Theta}_k$, the stacked profile (Equation~\eqref{eq:theta_hat_stacking}) on the remaining regions. The Jackknife covariance matrix element $\widehat{\mathbf{C}}^\jk_{ij}$ between radius $R_i$ and $R_j$ is given by
\begin{equation}
\begin{split}
    \widehat{\mathbf{C}}^\jk_{ij} =&
    \frac{N_\jk - 1}{N_\jk}
    \sum_{k = 1}^{N_\jk}
    \left(\pstackm[\jk]{i}- \pstack[k]{i}\right)\\
    &\times\left(\pstackm[\jk]{j}- \pstack[k]{j}\right)
\end{split}
\end{equation}
with the mean
\begin{equation}
    \pstackm[\jk]{i} = \frac{1}{N_\jk}\sum_{k = 1}^{N_\jk} \pstack[k]{i}.
\end{equation}
In \clmm, the split of the cluster catalog in $N_{\rm JK}$ independent and equal-sized regions relies on a HEALPix \citep{2005Gorskihealpix} subdivision. The free parameter is $N_{\rm side}$, linked to the number of subdivisions (pixels) of the three-dimensional sphere by $N_{\rm pix}=  12\times N_{\rm side}^2$. The precision in the Jackknife method to recover the true covariance depends on both the number of Jackknife sub-regions and the number of radial bins, and $N_\jk$ can be chosen to reach a given precision \cite[see e.g.][]{Friedrich2016cov}. \citet{Shirasaki2017covarianceDeltaSigma} has shown that the Jackknife method gives a reasonably accurate estimation of the true covariance for separation scales that are smaller than the size of Jackknife regions; however, it can underestimate the true error at the larger separations.

(ii) Another method is bootstrap resampling \citep[used e.g. in ][]{Parroni2017CLlensingcov}, computed as follows: for each stack, we create $N_\bs$ ensembles of clusters by selecting the same number of clusters but taking them with replacement. We compute $N_\bs$ times the stacked profile $\widehat{\Theta}_k$ for each bootstrap resampling, and we compute the covariance matrix as :
\begin{equation}
\begin{split}
    \widehat{\mathbf{C}}^\bs_{ij} =&
    \frac{1}{N_\bs - 1}\sum_{k = 1}^{N_\bs}
    \left(\pstackm[\bs]{i}- \pstack[k]{i}\right)\\
    &\times\left(\pstackm[\bs]{j}- \pstack[k]{j}\right)
\end{split}
\end{equation}
where
\begin{equation}
    \pstackm[\bs]{i} = \frac{1}{N_\bs}\sum_{k = 1}^{N_\bs} \pstack[k]{i}.
\end{equation}

(iii) The simplest is the \textit{sample} covariance matrix estimation, used in e.g. \citet{Simet2017SDSSmassrichness}. For a given stack of $N_{\rm cl}$ clusters,
we can use the individual profiles $\Theta_k$ to estimate the sample covariance as:

\begin{equation}
\begin{split}
    \widehat{\mathbf{C}}^\SC_{ij} =&
    \frac{1}{N_{\rm cl} (N_{\rm cl} - 1)}\sum_{k = 1}^{N_{\rm cl}}
    \left(\pstackm[\SC]{i}- \pradial[k]{i}\right)\\
    &\times\left(\pstackm[\SC]{j}- \pradial[k]{j}\right),
\end{split}
\end{equation}
where
\begin{equation}
    \pstackm[\SC]{i} = \frac{1}{N_{\rm cl}}\sum_{k = 1}^{N_{\rm cl}} \pradial[k]{i}.
\end{equation}
These three different methods can be used in \clmm to estimate the cluster lensing stacked profile covariance matrix from data, and can be used to derive cluster properties such as the effective mass, the concentration, etc.
The methods are compared in Section~\ref{sec:stack_dc2}.

\section{Examples of application of \clmm\ on precursor data}
\label{sec:application_on_data}
This section illustrates the new functionalities of CLMM, developed in view of the analysis of Rubin data, by presenting its application on precursor data set.
Four case examples are shown, where \clmm is used to measure individual reduced shear profiles from HSC data (\S\ref{sec:hsc}) and DES data (\S\ref{sec:des}); stacked $\Delta \Sigma$ profiles from DESC DC2 data (\S\ref{sec:stack_dc2}); and triaxiality measurements for mock clusters (\S\ref{sec:triaxiality_mock}). We refer the reader to the corresponding notebooks provided with this code release for details on analysis choices and only provide a high-level summary below.
\subsection{Individual profile using HSC data}
\label{sec:hsc}

As a first example\footnote{More details are given in the \clmm\ demo notebook:
\url{Figure_08_Fit_Halo_mass_to_HSC_data_paper2.ipynb}
}, we applied \clmm to observationnal data from the Hyper Suprime-Cam Subaru Strategic Program (HSC SSP) third public data release~\citep{Aihara2018a,Aihara2018b,Aihara2019,Mandelbaum2018a,mandelbaum18}\footnote{\url{https://hsc-release.mtk.nao.ac.jp/doc/}}. This data set is particularly relevant as HSC SSP has comparable image quality, data processing pipelines, and data product formats to the Rubin LSST.

We focus on two relatively high-mass clusters at medium redshift as our targets, to maximise the lensing signal:
HWL16a-026 ($z_{\rm cl}=0.424,\,\alpha=130.5895,\,\delta=1.6473$)) and HWL16a-034 ($z_{\rm cl}=0.315,\, \alpha=139.0387,\, \delta=-0.3966$) from the catalog presented in ~\citet{Hamana2020}.

To build the corresponding source samples, we query
the HSC database\footnote{\url{https://hsc-release.mtk.nao.ac.jp/datasearch/helps/sql_search}}:
We use the \texttt{S16A} version of the shape catalog (with shapes measured by the HSM algorithm, \citealp{mandelbaum2005systematic})
and obtain redshifts by cross-matching with the \texttt{frankenz} photo-z catalog\footnote{\url{https://github.com/joshspeagle/frankenz}} \citep[Speagle et al. in prep.]{Tanaka2018}.
We use the quality cuts and shear calibration adopted in~\citet{Mandelbaum2018a}.
Finally,  we select galaxies around the cluster and with photometric redshift $z>z_{\rm cl} + 0.1$ (to remove foreground galaxies).

\begin{figure*}
    \includegraphics{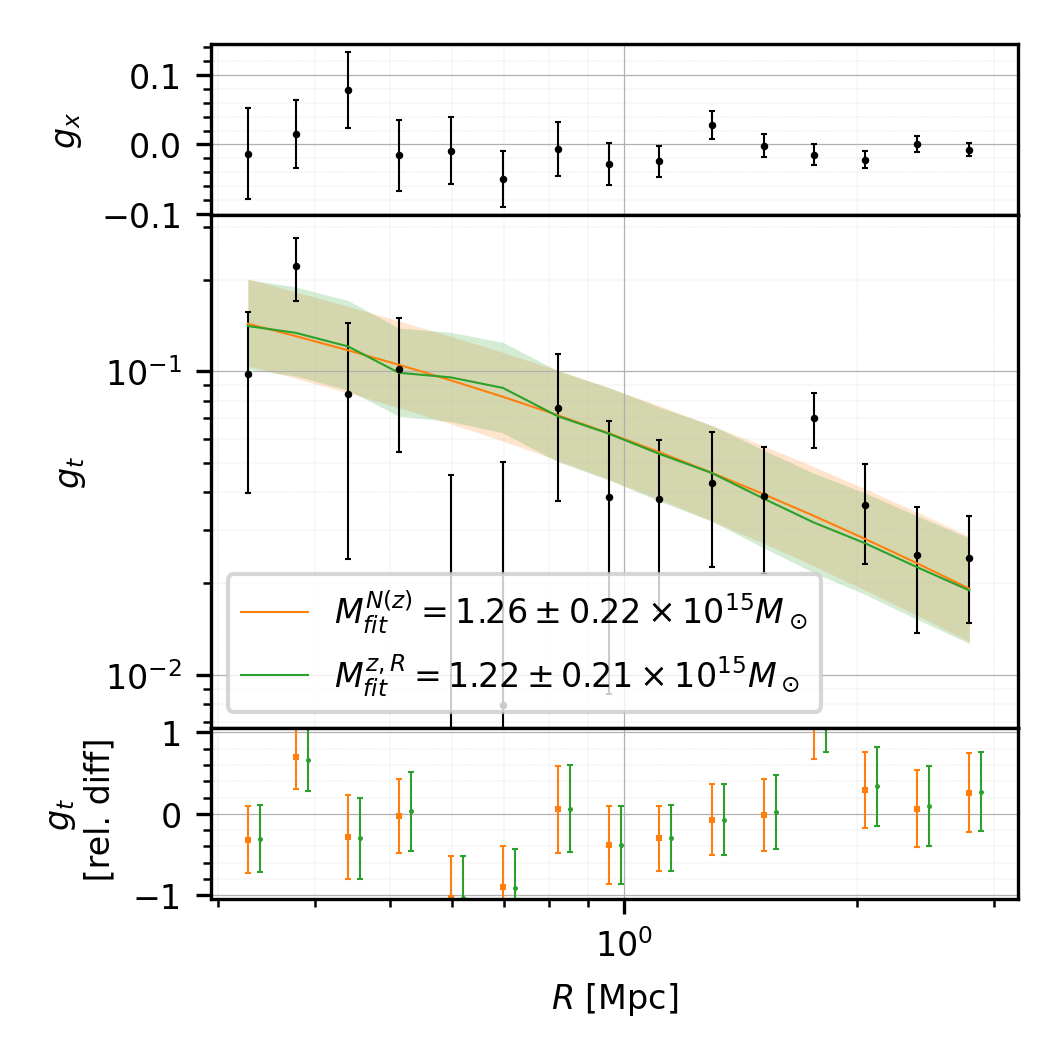}
    \includegraphics{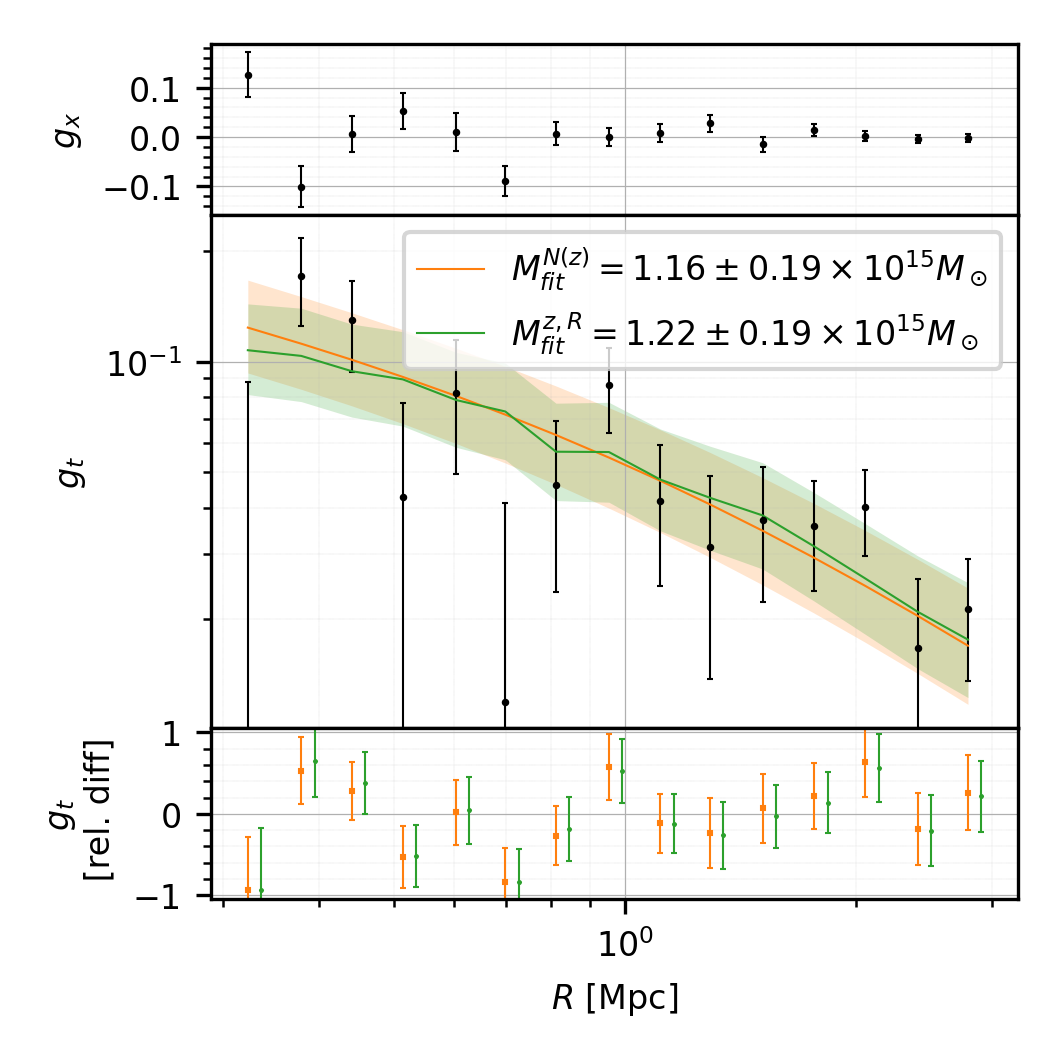}
    \caption{
Reduced shear profiles for two clusters, obtained by applying \clmm to HSC SSP data.
    The upper and middle panels show the cross and tangential components, respectively. The bottom panels show the relative differences between the data and model for the tangential component.
    The filled region gives $\pm3\sigma$ from the models. The model in orange is computed using the source redshift distribution, whereas the green one is evaluated using individual source redshifts.
    \textit{Left}: Cluster HWL16a-026 at $z=0.424$.
    \textit{Right}: Cluster HWL16a-034 (ACO 776) at $z=0.315$. }
    \label{fig:hsc_example}
\end{figure*}

The cluster information is stored in \clmm by creating a \texttt{GalaxyCluster} object, with the catalog of selected galaxies ingested as a \texttt{GCData} object.
We use methods from the \texttt{GalaxyCluster} object to
measure the reduced tangential and cross shear profiles
between 0.3 Mpc and 3 Mpc,
as displayed in Figure~\ref{fig:hsc_example}.
The cross signal (top panels) serves as a null test and is indeed consistent with zero.
The tangential profiles reach $g_t\sim0.1$ at $R\sim0.3$~Mpc, and decrease with radius as expected.

Figure~\ref{fig:hsc_example} also shows the
best fit models, obtained by fitting the cluster mass assuming a NFW parametrization with a concentration $c_{\rm 200c}=4$, either i) using the new CLMM functionality implemented from Eq.(\ref{eq:g_order1}) (i.e. assuming only the overall redshift distribution is known; orange, $M_{\rm fit}^{N(z)}$ in the figure), or ii) building a model using individual source redshifts and radial distances to compute the mean shear in each radial bin (green, $M_{\rm fit}^{z, R}$ in the figure). The bottom panels show the relative difference between the fitted models and the measured profiles. For the two clusters, the two methods yield compatible results.\footnote{These mass estimates are within $2\sigma$ of the ones reported in \citet{Hamana2020}. The methodologies and the cosmology used here differ from their analysis, which could explain the offset in the fitted values. However, our goal here is to illustrate an application of \clmm on real data rather than a detailed comparison, which is beyond the scope of this paper.}

\subsection{Individual profile using DES data}
\label{sec:des}
In the same spirit as the previous section, we provide a demonstration notebook presenting using Dark Energy Survey (DES) data with \clmm\footnote{\url{https://github.com/LSSTDESC/CLMM/tree/main/examples/Paper_v2.0/Fit_Halo_mass_to_DES_data_paper2.ipynb}}.
We select two high-richness clusters
from the DES Y1 \rdmp Catalog \footnote{\url{https://des.ncsa.illinois.edu/releases/y1a1/key-catalogs/key-redmapper}}
\citep{McClintock2019mass}: ID 1 (SPT-CL J0254-5856/5857;
at z=$0.430$) and ID 10 (MCS J2135.2-0102, 	PSZ2 G053.44-36.25;
at $z=0.326$).

The example notebook first demonstrates how to access the DES Y1 \metacal shape catalog~\citep{Zuntz2018} from NOIRLab Data Lab\footnote{\url{https://datalab.noirlab.edu/query.php?name=des_dr1.shape_metacal_riz_unblind}} for lensing analysis, and to cross match it with the DES DR1 photo-z catalog\footnote{\url{https://datalab.noirlab.edu/query.php?name=des_dr1.photo_z}}~\citep{Hoyle2018}.
In principle, the matching between the shear and photo-z catalogs can introduce a selection bias on the shear of up to 5\% \citep{2025MNRAS.543.4156Y}. However, we expect its effect on our results to be small given that the cluster shear is typically one order of magnitude higher than cosmic shear and, for a single cluster, the uncertainty caused by shape noise is much higher than other sources of uncertainty.

To build the source samples for the two clusters, we
require the photo-z parameters to obey $\texttt{minchi2}<1$, $\texttt{z\_sigma}<0.1$, and the shape measurement flag to be $\texttt{flags\_select}=0$, to ensure the quality of source galaxies.
The shape catalog gives the lensing shear information measured from the \metacal algorithm~\citep{metacal}, and the shear signal is estimated by the mean ellipticity of an ensemble of galaxies divided by a shear response.
As detailed in the notebook, we use a version of the calibration presented in \citet{McClintock2019mass} in this analysis\footnote{In particular, we ignore the selection bias of the shear response since it is typically at percent level.}.
Finally, we exclude sources with shear estimates larger than 1, and we select sources that have redshifts in the range $z_{\rm cl}+0.1<z<1.5$.

\begin{figure*}
    \centering
    \includegraphics[scale=.99]{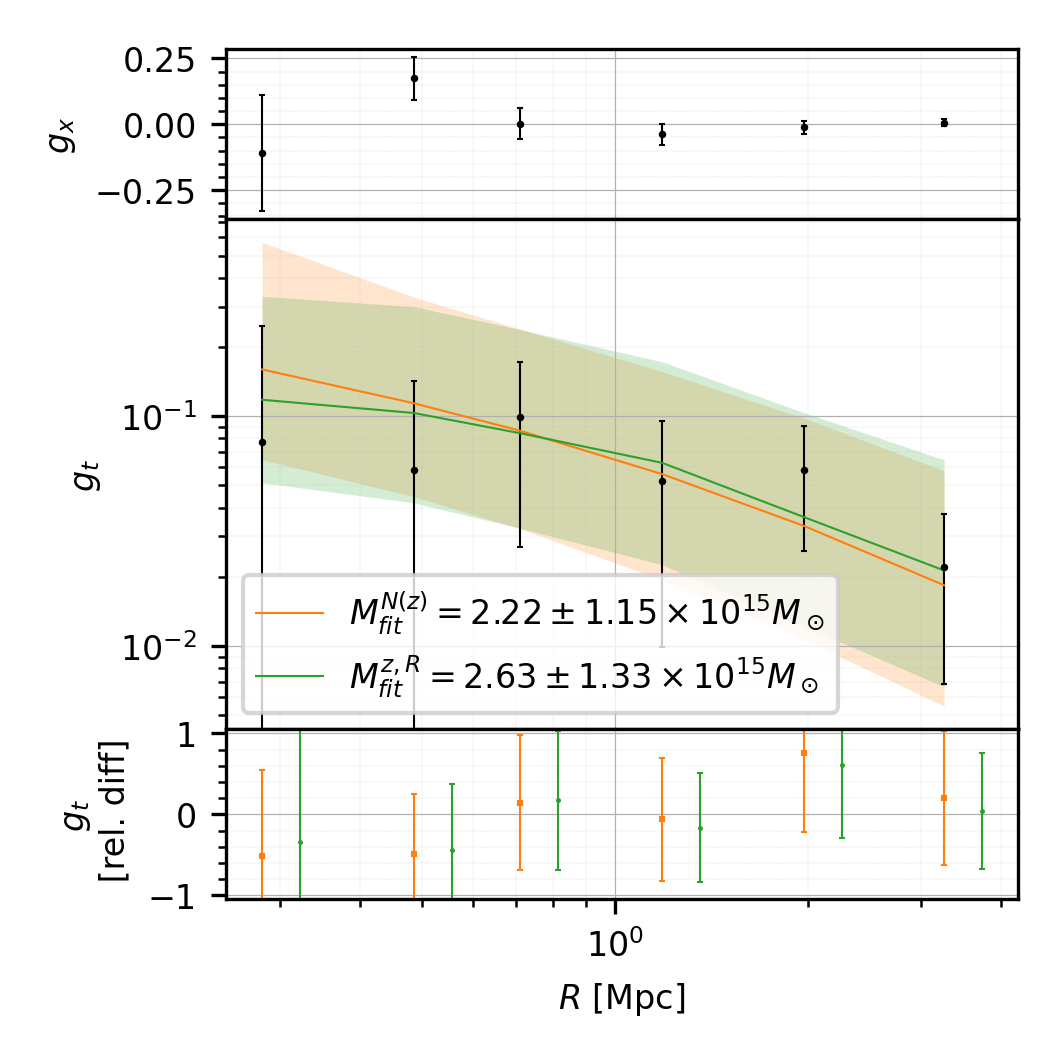}
    \includegraphics[scale=.99]{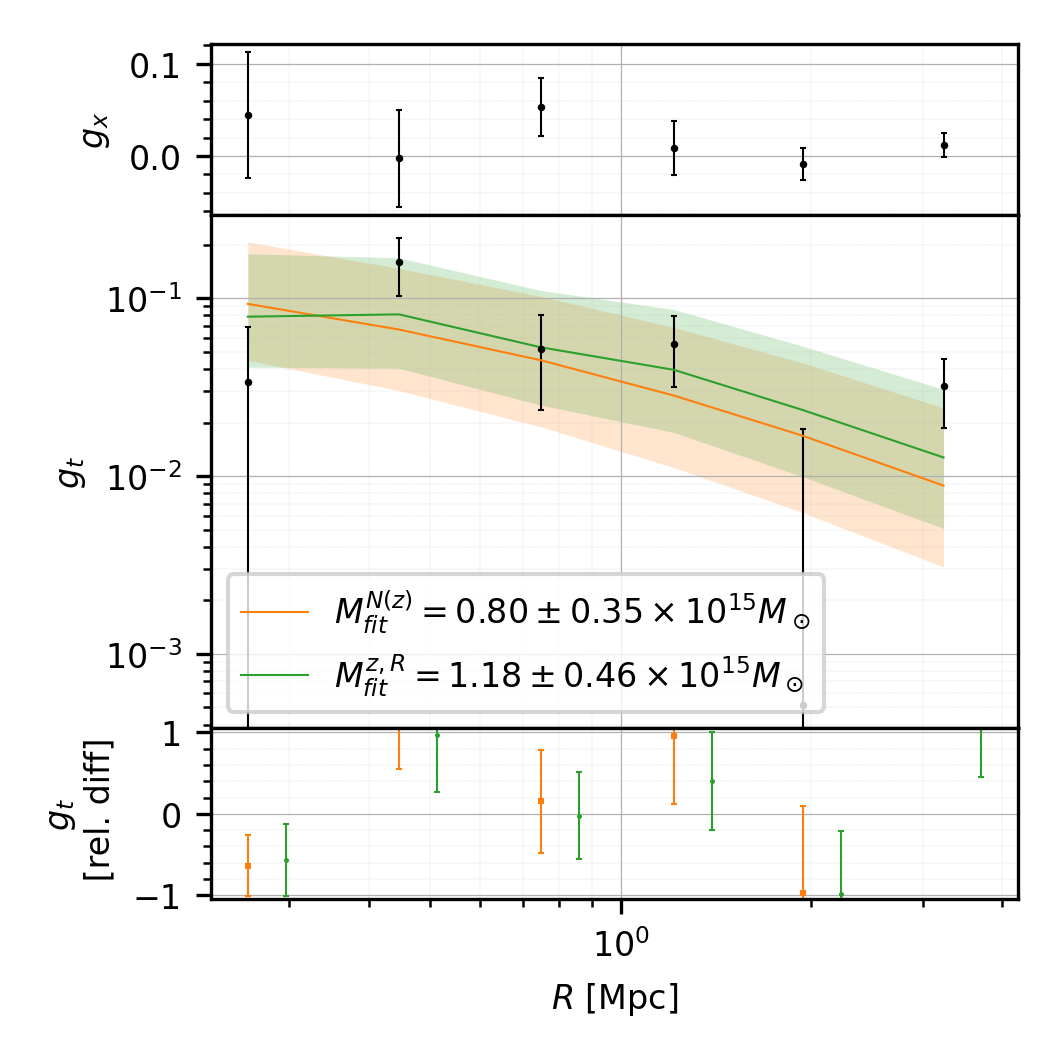}
    \caption{ Reduced shear profiles for two clusters, obtained by applying \clmm to DES data. The legend is the same as in Fig. \ref{fig:hsc_example}.
    \textit{Left}: Cluster with \rdmp ID 1 at $z=0.430$.
    \textit{Right}: Cluster with \rdmp ID 10 at $z=0.326$. }

    \label{fig:des_example}
\end{figure*}

From this calibrated source sample, we proceed as in Section \ref{sec:hsc},
to compute the tangential and cross shear profiles, and then fit the tangential profile using \clmm (using the same modeling as for the previous HSC example)\footnote{We now use 6 bins from 0.2 Mpc to 4 Mpc (compared to HSC SSP, DES has lower depth, and thus we use larger bins to increase the lensing S/N).}. The results are shown in Figure \ref{fig:des_example} and follow the same convention and notations as in the previous section.
For the two clusters, we see that the two methods yield compatible results, as quoted in the figure.\footnote{Note that these estimates are close to those reported in the literature~\citep[within $1\sigma$;][]{Williamson2011,Bocquet2019,Sereno2017,Planck2016}.}

\subsection{Stacked profiles example using the DESC DC2 data}
\label{sec:stack_dc2}

\begin{figure*}
    \centering
    \includegraphics[width=0.45\linewidth]{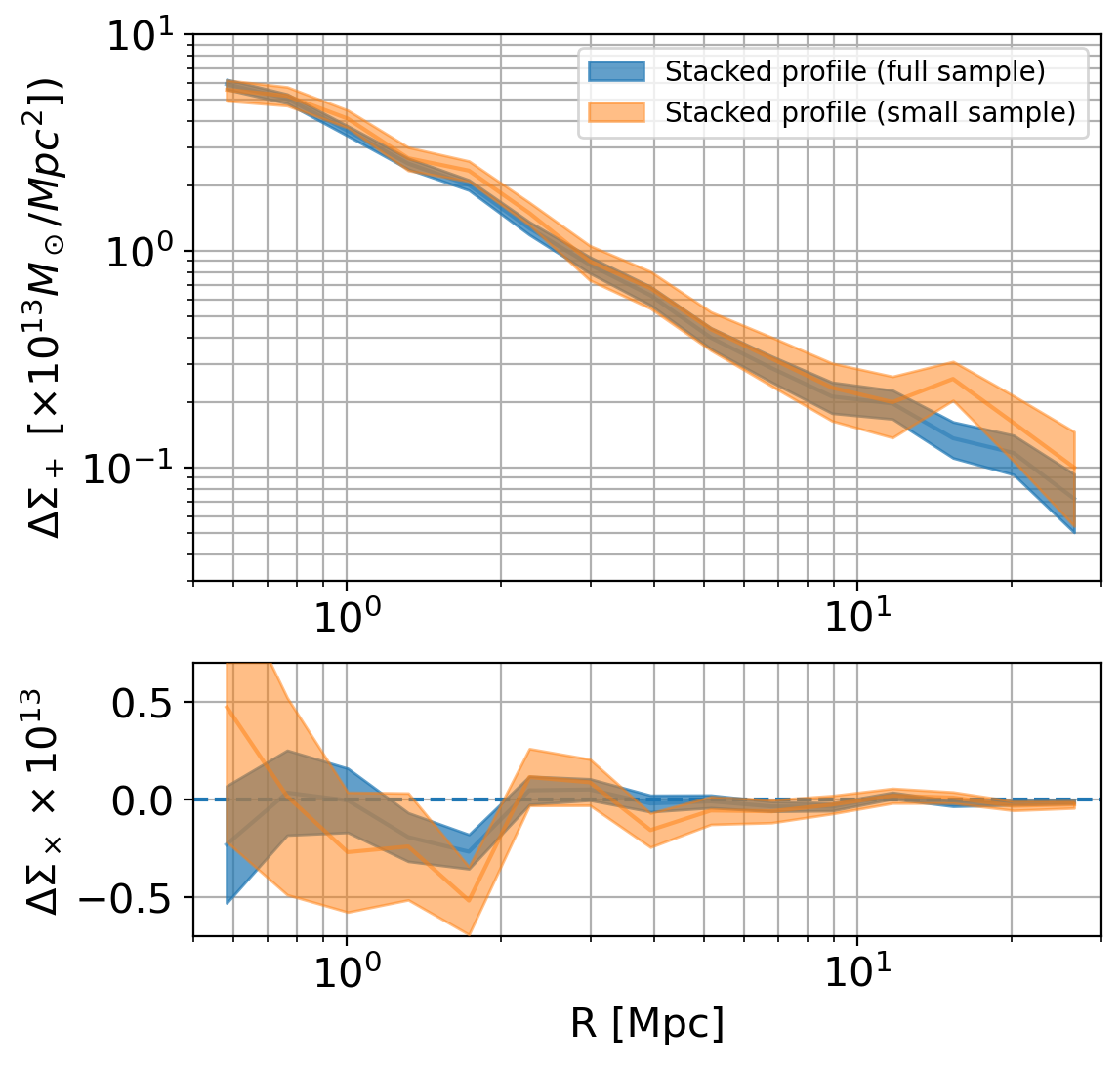}
    \includegraphics[width=0.5\linewidth]{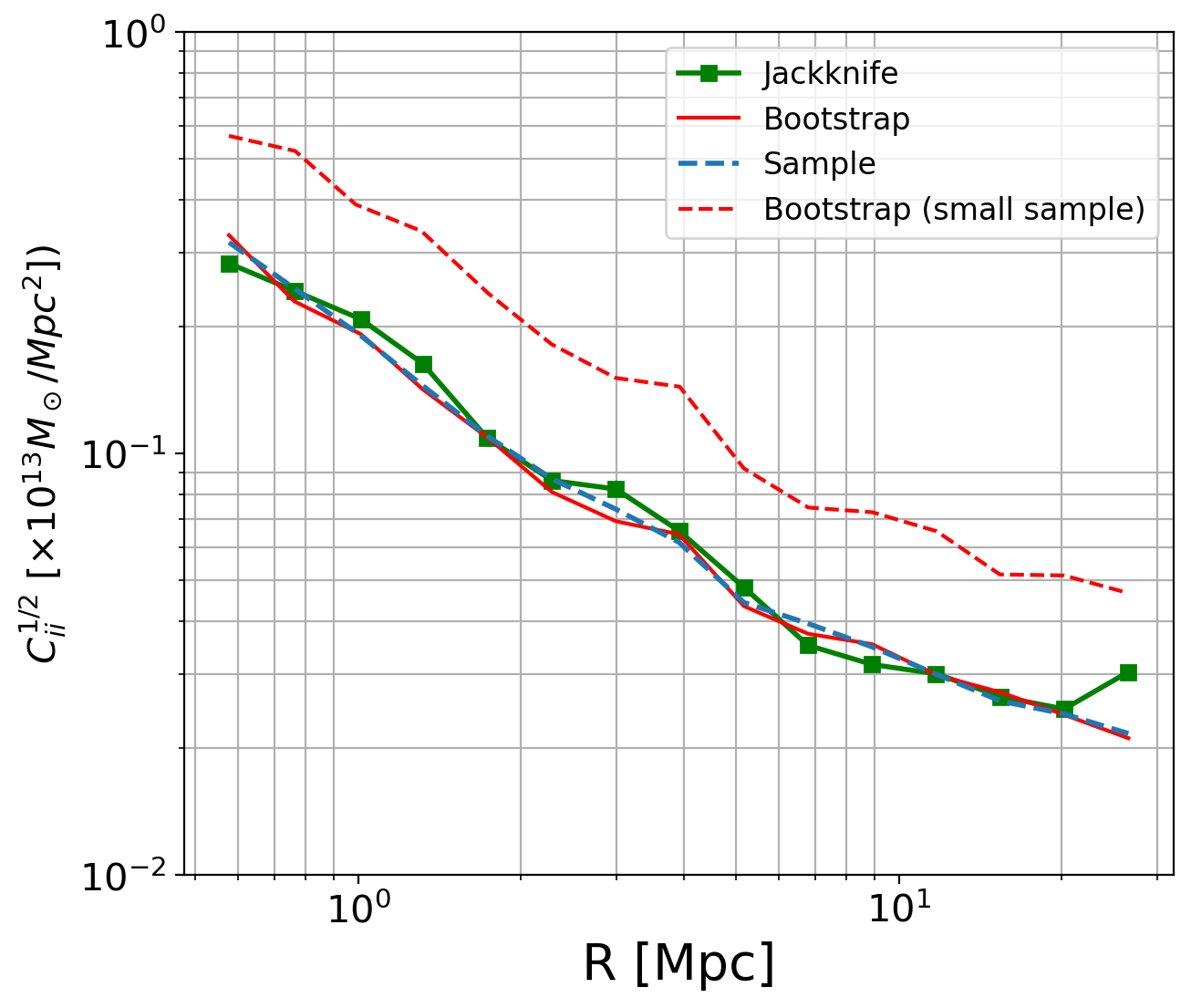}
    \caption{Left: Stacked excess surface density profiles (upper panel: tangential component, lower panel: cross component) for two different cluster samples (the full sample is represented in shaded blue, the reduced sample is in shaded orange). Error bars are obtained with the data-based Sample Covariance approach. Right: Diagonal element of the stacked profile covariance matrix (square root). The sample covariance approach is shown in dashed blue, the Bootstrap in green, and the Jackknife in green. The red dashed line shows the errors for the small cluster sample (Bootstrap-only).
    }
    \label{fig:DC2_cluster_ensemble}
\end{figure*}
\begin{figure}
    \centering
    \includegraphics[width=\columnwidth]{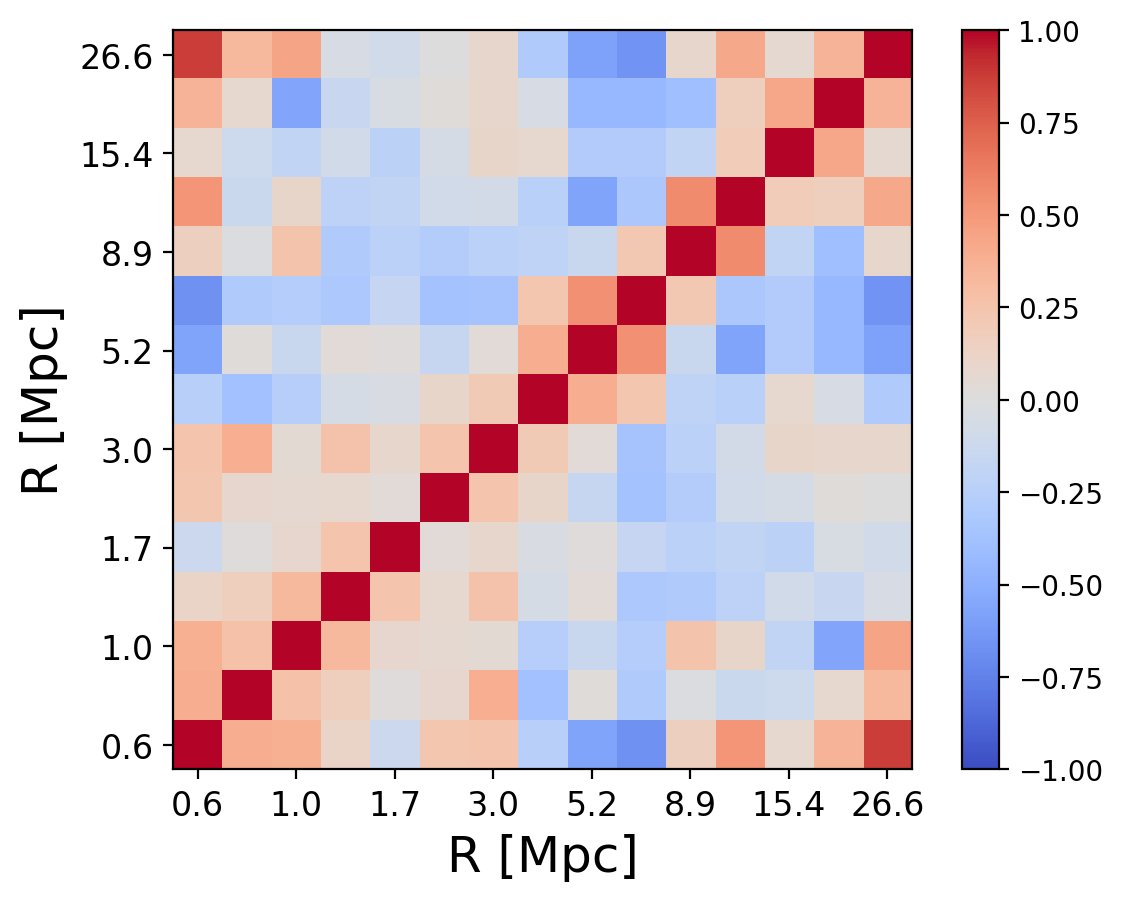}
    \caption{Correlation matrix of the stacked cluster tangential $\Delta\Sigma$ profile (for the full cluster sample), estimated using the Bootstrap method with $N_{\rm b} = 500$.
    }
    \label{fig:DC2_cluster_ensemble_corr}
\end{figure}

In this section, we illustrate the use of the \code{ClusterEnsemble} object using the cosmoDC2 extragalactic catalog \citep{korytov2019cosmodc2}  from the LSST DESC Data Challenge 2 \citep[DC2,][]{abolfathi2020lsst}.
For this example, we consider a restricted sample of $\sim 450$ \code{redMaPPer} clusters detected in cosmoDC2 with richness
$20 < \lambda < 40$ and redshift $0.2 < z < 0.4$. We follow the methodology in \citet{Payerne2025dc2CL} to select the source samples.

We first compute the individual profiles for each cluster, store them into a \code{ClusterEnsemble} object, and use the \code{make\_stacked\_radial\_profile} method to compute the stacked excess density profiles of the ensemble. The left panels in Fig.~\ref{fig:DC2_cluster_ensemble} show the tangential (top) and cross (bottom) components of the profile in blue. The error band corresponds to the diagonal elements of the sample covariance matrix obtained using the \code{compute\_sample\_covariance} method. Running the same analysis with a smaller sample of 100 clusters, randomly selected in the full sample, yields the compatible but more uncertain orange profiles, as expected.

The right panel in Fig.~\ref{fig:DC2_cluster_ensemble} compares the diagonal elements of the covariance matrices obtained using the \code{sample} (dashed blue), \code{bootstrap} (solid/dashed red, for the full/small sample) and \code{jackknife} (solid green with symbols) approaches\footnote{The bootstrap covariance calculation uses 500 samples. The Jackknife covariance relies on 32 equal-sized regions over the cluster catalog area.}. For the full sample, the three methods are in good agreement, however affected by the intrinsic noise of their respective data-based methods.
Overall, the error decreases with the radius, as the number of galaxies in a given radial bin increases. The errors of the small cluster sample (dashed red) are larger by a factor $\sqrt{N_{\rm tot}/N_{\rm small}} \sim 2.1$ compared to the full stack, as expected.
Below $R\sim$ 5 Mpc, the stacked profile error follows the simple relation $C_{ii}^{1/2} \propto 1/\sqrt{N_{\rm gal}(R)} \propto R^{-1}$, from Poisson noise. At a radius of $R\sim$ 5 Mpc, the contribution from the correlated large-scale structure becomes increasingly important.

 For illustration, we show in Figure \ref{fig:DC2_cluster_ensemble_corr} the correlation matrix of the stacked cluster $\Delta\Sigma$ profile obtained from the Bootstrap covariance matrix of the full cluster sample,  highlighting the (anti-)correlations between the various radial bins of the stacked profile.

\subsection{Cluster triaxiality analysis using mock data}
\label{sec:triaxiality_mock}

To evaluate the performance of \clmm's triaxiality measurements and theoretical prediction, we produced a mock catalog of elliptical halos based on the methodology of \cite{2026arXiv260506587F}.
We considered a halo with the median characteristics of the DES Y1 sample: a mass $M_{\rm 200c}=2\times10^{14}M_\odot$ (corresponding to a \rdmp richness $\lambda\sim30$), a redshift $z=0.47$, and a concentration $c_{\rm 200c}\sim4$ \citep[following the relation in ][]{2018ApJ...859...55C}.
The axis ratio was set to $b/a=2/3$, which matches the mean shape of the ~\citet{2024MNRAS.530.2251F} observed sample.
The sources are simulated at a redshift plane $z=0.8$, with a density of 2.4k per arcmin$^2$, which virtually corresponds to a stack of $\sim200$ Y1 LSST clusters.

\begin{figure*}[htbp]
    \centering
    \includegraphics{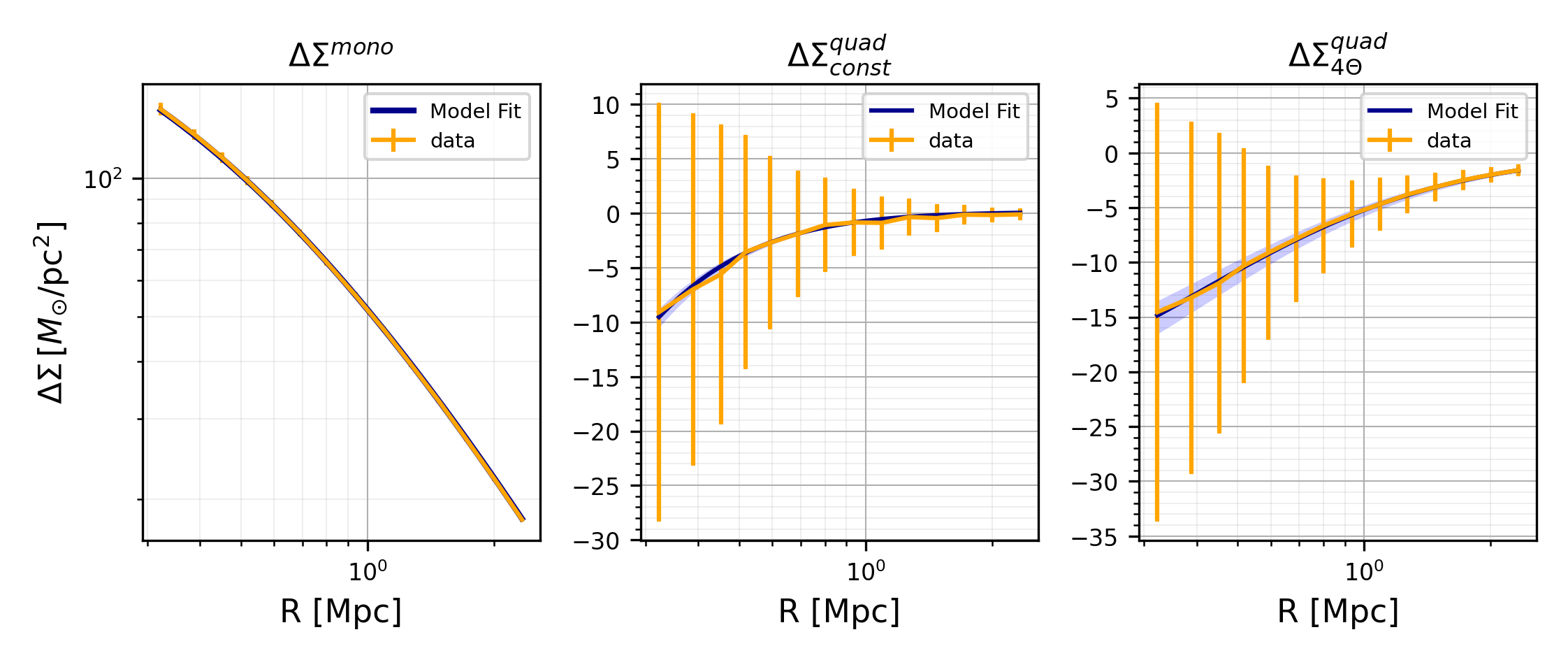}
    \caption{Excess surface density profiles for a mock halo with $q = 2/3$. The left panel shows the profile coming from the monopole (Equation~\eqref{eq:triaxiality_monopole}), while the middle and right panels show the profiles from two estimators of the quadrupole component (Equations ~\eqref{eq:triaxiality_const} and ~\eqref{eq:triaxiality_4theta}). The data points are indicated in orange, while the fit to the data is shown in blue.
    }
    \label{fig:triaxiality_fit}
\end{figure*}

We use the \code{compute\_tangential\_and\_cross\_components} function with the \code{include\_quadrupole} flag set to \code{True} to measure the different triaxial components of the weak lensing profile. The model to fit the mass and concentration of the mock cluster is then evaluated using the function \code{compute\_excess\_surface\_density\_triaxial}.
The results are shown in Figure~\ref{fig:triaxiality_fit}, in which the input parameters of the elliptical halo are well recovered: $M_{\rm 200c}=2.01\pm0.01\times10^{14}M_\odot$, $q=0.69\pm0.03$.

\section{Summary and perspectives}
\label{sec:summary}

In the current context of observational cosmology with big data, stable, efficient, reliable, and transparent software becomes more important to ensure the quality of the scientific results. In this context, we present the v2.0 release of \clmm.

    This library was developed to provide both a complete standalone tool to be used by the cluster community at large, and as an essential part of the larger DESC pipeline for cluster cosmology. The version released with this paper was designed to contain the necessary functionalities to perform the variety of DESC cluster weak lensing-related analyses anticipated with the first Rubin data.

    Compared to the first version of the code, two major new functionalities concern the ability to perform stacked weak lensing analyses, as well as the option to go beyond the spherically symmetric assumption and use multipoles of the weak lensing signal to account for cluster triaxiality.

    These are supplemented by a series of smaller, yet important, updates to the v1 functionalities, including: implementation of the two-halo term that becomes dominant on large scales; boost factor computation to account for possible source sample contamination; correction due to cluster miscentering; modeling of the magnification bias, as an alternative to the shear; flexibility in how redshift information is handled and approximate shear profile predictions when individual source redshifts are not known.

With the imminent arrival of LSST data, we also augment this release with a set of new demonstration notebooks showing how to use \clmm{} with Stage~III datasets from DES and HSC, but also how to apply the new stacked shear and triaxiality functionalities.

Looking forward to the first DESC cluster cosmology analyses, \clmm{} is currently being integrated with other DESC tools into an end-to-end cluster cosmology pipeline. Describing this pipeline goes beyond the scope of this paper, and it will be presented in further work; we nonetheless refer the interested reader to Appendix \ref{sec:interface} for a broad overview of how \clmm{} is being interfaced in this landscape.

\subsection*{Data availability}
The data used in this paper come either from publicly available sources or were generated by the {\tt mock\_data} module of the \clmm code, all of which can be accessed via the notebooks provided in the \clmm\ repository at
\url{https://github.com/LSSTDESC/CLMM/tree/main/examples/Paper_v2.0}.
Auxiliary documentation, containing a full description of the code API and several examples of its usage, is available at \url{https://lsstdesc.org/CLMM}.

\subsection*{Authors' Contribution}
\input{contributions}

\subsection*{Acknowledgments}

This paper has undergone internal review in the LSST Dark Energy Science Collaboration.
The authors would like to thank Tomomi Sunayama and Joe Zuntz for serving as the LSST-DESC publication review committee whose comments and suggestions that improved the quality of this manuscript.

\input{desc-tex/ack/standard}

The Hyper Suprime-Cam (HSC) collaboration includes the astronomical communities of Japan and Taiwan, and Princeton University. The HSC instrumentation and software were developed by the National Astronomical Observatory of Japan (NAOJ), the Kavli Institute for the Physics and Mathematics of the Universe (Kavli IPMU), the University of Tokyo, the High Energy Accelerator Research Organization (KEK), the Academia Sinica Institute for Astronomy and Astrophysics in Taiwan (ASIAA), and Princeton University. Funding was provided by the FIRST program from the Japanese Cabinet Office, the Ministry of Education, Culture, Sports, Science and Technology (MEXT), the Japan Society for the Promotion of Science (JSPS), Japan Science and Technology Agency (JST), the Toray Science Foundation, NAOJ, Kavli IPMU, KEK, ASIAA, and Princeton University.

This paper makes use of software developed for Vera C. Rubin Observatory. We thank the Rubin Observatory for making their code available as free software at http://pipelines.lsst.io/.

This paper is based on data collected at the Subaru Telescope and retrieved from the HSC data archive system, which is operated by the Subaru Telescope and Astronomy Data Center (ADC) at NAOJ. Data analysis was in part carried out with the cooperation of the Center for Computational Astrophysics (CfCA), NAOJ. We are honored and grateful for the opportunity of observing the Universe from Maunakea, which has the cultural, historical and natural significance in Hawaii.

This project used public archival data from the Dark Energy Survey (DES). Funding for the DES Projects has been provided by the U.S. Department of Energy, the U.S. National Science Foundation, the Ministry of Science and Education of Spain, the Science and Technology FacilitiesCouncil of the United Kingdom, the Higher Education Funding Council for England, the National Center for Supercomputing Applications at the University of Illinois at Urbana-Champaign, the Kavli Institute of Cosmological Physics at the University of Chicago, the Center for Cosmology and Astro-Particle Physics at the Ohio State University, the Mitchell Institute for Fundamental Physics and Astronomy at Texas A\&M University, Financiadora de Estudos e Projetos, Funda{\c c}{\~a}o Carlos Chagas Filho de Amparo {\`a} Pesquisa do Estado do Rio de Janeiro, Conselho Nacional de Desenvolvimento Cient{\'i}fico e Tecnol{\'o}gico and the Minist{\'e}rio da Ci{\^e}ncia, Tecnologia e Inova{\c c}{\~a}o, the Deutsche Forschungsgemeinschaft, and the Collaborating Institutions in the Dark Energy Survey.

The Collaborating Institutions are Argonne National Laboratory, the University of California at Santa Cruz, the University of Cambridge, Centro de Investigaciones Energ{\'e}ticas, Medioambientales y Tecnol{\'o}gicas-Madrid, the University of Chicago, University College London, the DES-Brazil Consortium, the University of Edinburgh, the Eidgen{\"o}ssische Technische Hochschule (ETH) Z{\"u}rich,  Fermi National Accelerator Laboratory, the University of Illinois at Urbana-Champaign, the Institut de Ci{\`e}ncies de l'Espai (IEEC/CSIC), the Institut de F{\'i}sica d'Altes Energies, Lawrence Berkeley National Laboratory, the Ludwig-Maximilians Universit{\"a}t M{\"u}nchen and the associated Excellence Cluster Universe, the University of Michigan, the National Optical Astronomy Observatory, the University of Nottingham, The Ohio State University, the OzDES Membership Consortium, the University of Pennsylvania, the University of Portsmouth, SLAC National Accelerator Laboratory, Stanford University, the University of Sussex, and Texas A\&M University.

Based in part on observations at Cerro Tololo Inter-American Observatory, National Optical Astronomy Observatory, which is operated by the Association of Universities for Research in Astronomy (AURA) under a cooperative agreement with the National Science Foundation.

\bibliographystyle{mnras}
\bibliography{main}

\appendix
\section{Interfacing with other \desc\ tools}
\label{sec:interface}

As pointed out in \paperI\, \clmm\ can be used as a standalone library open to the community for cluster WL profile measurements, but is also part of the \desc\ effort to produce robust software infrastructure enabling flexible multi-probe cosmology analyses. \clmm\ is, in particular, a key element of the DESC Cluster Cosmology pipeline that is currently under development.
As an example, here we discuss the infrastructure required for analyzing cluster counts and stacked $\Delta \Sigma$ profiles in bins of redshift and richness. Note that this configuration is only one possibility, and open to evolution.

The current configuration of the DESC Cluster Cosmology pipeline is presented in detail in Barroso et al. in prep, and we just give a summary below to highlight the role of \clmm in it.
The analysis begins with a cluster catalog produced by one of the cluster finder algorithms validated within DESC, along with a source galaxy catalog and some configuration metadata such as redshift and richness binning information. The data-side computations are performed within \code{TXPipe}, which includes the measurement of stacked weak lensing  $\Delta \Sigma$ signals and cluster counts in redshift and richness bins. The profile measurements are performed internally by the \clmm\ library, which is being called from \code{TXPipe}. Additionally, \code{TXPipe} computes the empirical covariance of the $\Delta \Sigma$ stacked profiles, again using \clmm.

On the theoretical side, model predictions for the observables are generated in the \code{CROW} library, which uses the theory functionalities from \clmm\ to compute the model for the weak lensing signal.
The theoretical covariance for cluster counts (which has the same binning scheme used in the data) is computed by \code{TJPCov}.
These components are then fed into \code{Firecrown}, the code responsible for the DESC likelihood framework, where the final statistical inference is performed using an external parameter sampler.

\clmm\ has been structured to provide flexibility and efficiency within the DESC cluster cosmology pipeline.
Hence, it can also be easily applied to different configuration of the cluster pipeline, e.g. using either individual reduced shear profiles, or  weak lensing masses as data vector. Additionally, the object-oriented theory interface of \clmm\ allows for a better integration with the theory prediction module and, consequently, faster computations.
Finally, most DESC cluster modules were set up to use \ccl\ for cosmological calculations, thereby enabling consistent integration across codes. Additionally, \clmm\ also provides functions to ensure the cosmology used at different points of the data processing is consistent.

\vspace{10cm}

\end{document}

%% file: diagram.tex
\usetikzlibrary{arrows.meta}
\usetikzlibrary{shapes.geometric}
\usetikzlibrary{shapes.misc}
\usetikzlibrary{positioning}
\definecolor{TheoryColor}{HTML}{38761d}
\definecolor{DataOpsColor}{HTML}{1155cc}
\definecolor{GCColor}{HTML}{85200c}
\definecolor{CosmoColor}{HTML}{351c75}
\definecolor{UtilsColor}{HTML}{0c433d}
\definecolor{GCDataColor}{HTML}{7f6000}
\definecolor{SupportColor}{HTML}{000000}
\definecolor{EnsembleColor}{HTML}{440d32}

\definecolor{TheoryColorfill}{HTML}{63d297}
\definecolor{DataOpsColorfill}{HTML}{c9daf8}
\definecolor{GCColorfill}{HTML}{f4cccc}
\definecolor{CosmoColorfill}{HTML}{d9d2e9}
\definecolor{UtilsColorfill}{HTML}{d9ead3}
\definecolor{GCDataColorfill}{HTML}{fff2cc}
\definecolor{SupportColorfill}{HTML}{cccccc}
\definecolor{EnsembleColorfill}{HTML}{f9c0e6}

\definecolor{faintred}{HTML}{ea9999}
\definecolor{PipelineColor}{HTML}{85200c}
\definecolor{PipelineColorfill}{HTML}{f4cccc}
\definecolor{DataProductColor}{HTML}{38761d}
\definecolor{DataProductColorfill}{HTML}{63d297}
\tikzset{%
  >={Latex[width=2.5mm,length=2mm]},
    package/.style = {rectangle, draw=black,
        minimum width=5cm, minimum height=2.7cm,
        text width=4.8cm,
        text centered,
        line width=1.5pt,
        font=\sffamily},
    module/.style = {package, rounded corners},
    pipe/.style = {
        draw=PipelineColor,
        fill=PipelineColorfill,
        minimum width=4cm, minimum height=1cm,
        text width=3.8cm,
        text centered,
        line width=1.5pt,
        font=\sffamily},
    pipeline/.style = {rectangle, pipe},
    dataproduct/.style = {
        rounded corners,
        pipe,
        minimum width=.5cm,
        text width=3.cm,
        draw=DataProductColor,
        fill=DataProductColorfill,
        },
}
\newcommand{\diagText}[1]{\texttt{#1}}
\newcommand{\diagTitle}[3][black]{
    \scalebox{1.4}{
        \textbf{
            \color{#1}{#2}
        }
    }
}

\def\DataOps{{\color{DataOpsColor}\textbf{DataOps}}}
\def\GCData{{\color{GCDataColor}\textbf{GCData}}}
\newcommand{\diagram}{
\begin{tikzpicture}[
    node distance=3cm,
    every node/.style={fill=white, font=\sffamily},
    align=center
]
    \node(Utils)[
        package,
        draw=UtilsColor,
        fill=UtilsColorfill
    ]{
        \diagTitle[UtilsColor]{Utils}{} \\
        \diagText{Auxiliary functions (binning, conversions, simple computations)}
    };
    \node(DataOps)[
        package,
        above left=0.5cm and 1cm of Utils,
        draw=DataOpsColor,
        fill=DataOpsColorfill
    ]{
        \diagTitle[DataOpsColor]{DataOps}{} \\
        \diagText{Compute desired quantities from measurements (shear components, profiles)}
    };
    \node(GCData)[
        module,
        above=0.5cm of Utils,
        draw=GCDataColor, fill=GCDataColorfill
    ]{
        \diagTitle[GCDataColor]{GCData}{} \\
        \diagText{Astropy table with added functionality to check cosmology signature}
    };
    \node(GC)[
    module,
        below left=0.5cm and 1cm of Utils,
        draw=GCColor,
        fill=GCColorfill
    ]{
        \diagTitle[GCColor]{GalaxyCluster}{} \\
        \diagText{Cluster information (ID, ra, dec, z), \DataOps\ functionality, galaxy source catalog (\GCData), shear profiles (\GCData)}
    };
    \node(Ensemble)[
    module,
        left=1cm of Utils,
        draw=EnsembleColor,
        fill=EnsembleColorfill
    ]{
        \diagTitle[EnsembleColor]{ClusterEnsemble}{} \\
        \diagText{
        Ensemble containing clusters information and profiles, with stacking function 
        (weights computation, stacking, covariances estimation)
        }
    };
    \node(Cosmo)[
        package,
        below right=-1cm and 1cm of Utils,
        draw=CosmoColor,
        fill=CosmoColorfill
    ]{
        \diagTitle[CosmoColor]{Cosmology}{} \\
        (astropy, CCL, NumCosmo)\\
        \diagText{Cosmological operations\ (distances, parameter evolution)}
    };
    \node(Theory)[
        package,
        below=0.5cm of Utils,
        draw=TheoryColor,
        fill=TheoryColorfill
    ]{
        \diagTitle[TheoryColor]{Theory}{} \\
        (cluster\_toolkit, CCL, NumCosmo)\\
        \diagText{Theoretical prediction for lensing properties (surface density, shear, magnification)}
    };
    \node(Support)[
        package,
        above right=-1cm and 1cm of Utils,
        draw=SupportColor,
        fill=SupportColorfill
    ]{
        \diagTitle[SupportColor]{Support}{} \\
        \diagText{
        \hypersetup{linkcolor=.}
        Operations to test and facilitate the use of clmm (\hyperref[sec:mock_data]{mock data},  \hyperref[sec:samplers]{parameters fitting/sampling})}
    };
    
    \draw[line width=2pt, color=red, ->](Utils) -- (Support);
    \draw[line width=2pt, color=red, ->](GCData) -- (Support);
    \draw[line width=2pt, color=red, ->](Utils) -- (DataOps);
    \draw[line width=2pt, color=red, ->](Utils) -- (GCData);
    \draw[line width=2pt, color=red, ->](DataOps) -- (Ensemble);
    \draw[line width=2pt, color=red, ->](Theory) -- (GC);
    \draw[line width=2pt, color=red, ->](GCData) -- (DataOps);
    \draw[line width=2pt, dashed, color=faintred, ->](Cosmo) -- (Theory);
    \draw[line width=2pt, dashed, color=faintred, ->] plot [smooth, tension=.5] coordinates {(Cosmo.east) (8.5, 4.5) (-4, 5.5) (DataOps.north)};
    \draw[line width=2pt, color=red, ->](Utils) -- (Cosmo);
    \draw[line width=2pt, color=red, ->](Utils) -- (GC);
    \draw[line width=2pt, color=red, ->](Utils) -- (Theory);
    \draw[line width=2pt, color=red, ->](Utils) -- (Ensemble);
   \draw[line width=2pt, color=red, ->](DataOps.west) .. controls +(left:1cm) and +(left:1cm) .. (GC.west);
    \draw [line width=2pt, color=red, ->] plot [smooth, tension=.5] coordinates {
    (Theory.east) (5,-4) (9,-3.5) (9.2,0) (Support.east)
    };
    \end{tikzpicture}
}

%% file: authors.tex
% Author list file generated with: mkauthlist 1.2.4 
% mkauthlist -j tex -f -c LSST Dark Energy Science Collaboration --journal mnras --cntrb contributions.tex authors2.csv authors2.tex 
% Note: to add authors:
% 1. download the Overleaf project locally
% 2. add info to authors.csv
% 3. execute "make"
% 4. upload authors.tex to the Overleaf, replacing the previous version
\author
%[LSST Dark Energy Science Collaboration]
  {
  \parbox{\textwidth}{
    %\Large
    The CLMM Team$^{*}$:
    \textbf{M.~Aguena\orcidlink{0000-0001-5679-6747}}$^{1,2}$,
    C.~Combet\orcidlink{0000-0001-6487-1866}$^{3}$,
    S.~Fu\orcidlink{0000-0001-5422-1958}$^{4,5}$,
    C.~Payerne\orcidlink{0000-0002-1818-929X}$^{6}$,
    \textbf{M.~Ricci\orcidlink{0000-0002-3645-9652}}$^{7}$
    ~\\
    C.~Avestruz\orcidlink{0000-0001-8868-0810}$^{8,9}$,
    E. J.~Barroso\orcidlink{0000-0002-7988-9249}$^{10}$,
    C. L.~de Oliveira\orcidlink{0009-0005-8092-5496}$^{11}$,
    C.~Zhou\orcidlink{0000-0002-2897-6326}$^{12}$,
    H.~Fan\orcidlink{0009-0009-9294-7448}$^{13,14}$,
    C.~Murray\orcidlink{0000-0002-4668-1273}$^{15}$,
    A.~Niemiec\orcidlink{0000-0003-3791-2647}$^{3}$,
    N.~\v{S}ar\v{c}evi\'{c}\orcidlink{0000-0001-7301-6415}$^{16}$,
    T.~Shin\orcidlink{0000-0002-6389-5409}$^{17}$,
    R.~Solomon\orcidlink{}$^{18}$,
    R.~Srinivasan\orcidlink{0000-0002-4508-4581}$^{19}$,
    T.~Sunayama\orcidlink{}$^{20}$,
    Y.~Zhang\orcidlink{0000-0001-5969-4631}$^{21}$,
    J.~Zuntz\orcidlink{0000-0001-9789-9646}$^{22}$
    and The LSST Dark Energy Science Collaboration
}
  \vspace{0.4cm}
  \\
  \parbox{\textwidth}{
    % \tiny
    \scriptsize
    $^{*}$ {\textbf{Corresponding authors : 
    Michel Aguena (\href{mailto:aguena@inaf.it}{aguena@inaf.it}) and Marina Ricci (\href{mailto:ricci@apc.in2p3.fr}{ricci@apc.in2p3.fr})}} \\
    ~ \\
    $^{1}$ INAF-Osservatorio Astronomico di Trieste, via G. B. Tiepolo 11, I-34143 Trieste, Italy \\
    $^{2}$  Laboratório Interinstitucional de e-Astronomia - LIneA, Av. Pastor Martin Luther King Jr, 126 Del Castilho, Nova América Offices, Torre 3000/sala 817 CEP: 20765-000, Brazil \\
    $^{3}$ Université Grenoble Alpes, CNRS, IN2P3, LPSC, 38000 Grenoble, France \\
    $^{4}$ Kavli Institute for Particle Astrophysics and Cosmology \\
    $^{5}$ SLAC National Accelerator Laboratory, Menlo Park, CA, 94025, USA \\
    $^{6}$ Université Paris-Saclay, CEA, IRFU, 91191, Gif-sur-Yvette, France \\
    $^{7}$ Université Paris Cité, CNRS, APC-IN2P3, 75013 Paris, France  \\
    $^{8}$ University of Michigan, Ann Arbor, MI, USA \\
    $^{9}$ Leinweber Center for Theoretical Physics, Ann Arbor, MI, USA \\
    $^{10}$ LAPP, Université de Savoie, CNRS/IN2P3, Annecy-Le-Vieux, France \\
    $^{11}$ Departamento de Física, Universidade Estadual de Londrina, Rod. Celso Garcia Cid, Km 380, 86057-970, Londrina, Paraná, Brazil \\
    $^{12}$ Physics Department, University of California, Santa Cruz, CA 95064, USA \\
    $^{13}$ Institute of Astronomy and Astrophysics, Academia Sinica, Taipei 106319, Taiwan \\
    $^{14}$ Graduate Institute of Astrophysics, National Taiwan University, Taipei 106319, Taiwan \\
    $^{15}$ Université Paris-Saclay, Université Paris Cité, CEA, CNRS, AIM, 91191 Gif-sur-Yvette, France \\
    $^{16}$ Department of Physics, Duke University, Science Dr, Durham, NC 27710, USA \\
    $^{17}$ Carnegie Mellon University, Pittsburgh, PA, USA \\
    $^{18}$ Laboratoire d'Annecy De Physique Des Particules, CNRS, LAPP-IN2P3, 9 Chem. de Bellevue, 74940 Annecy, France \\
    $^{19}$ Stony Brook University, Stony Brook, NY, 11794, USA \\
    $^{20}$ Academia Sinica Institute of Astronomy and Astrophysics, No.1, Sec. 4, Roosevelt Rd, Taipei 106319, Taiwan, R.O.C \\
    $^{21}$ NSF NOIRLab, Tucson, AZ, USA \\
    $^{22}$ Institute for Astronomy, University of Edinburgh, Edinburgh EH9 3HJ, UK
  }
}

%% file: contributions.tex
% Macro generated by XXX
Author contributions are listed below : \\
M.~Aguena:
    conceptualization;
    data curation;
    methodology development;
    project administration;
    software;
    supervision;
    validation;
    visualization;
    writing - original draft;
    writing - review \& editing;
    DESC Builder.\\
C.~Combet:
    conceptualization;
    methodology development;
    project administration;
    software;
    writing - original draft;
    writing - review \& editing.\\
S.~Fu:
    methodology development;
    software;
    writing - original draft;
    writing - review \& editing.\\
C.~Payerne:
    methodology development;
    software;
    writing - review \& editing.\\
M.~Ricci:
    conceptualization;
    project administration;
    software;
    validation;
    writing - review \& editing;
    DESC Builder.\\
C.~Avestruz:
    software;
    DESC Builder.\\
E. J.~Barroso:
    software;
    writing - review \& editing.\\
C. L.~de Oliveira:
    software;
    writing - original draft.\\
C.~Zhou:
    validation.\\
H.~Fan:
    methodology development;
    software;
    writing - original draft;
    writing - review \& editing.\\
C.~Murray:
    methodology development;
    software.\\
A.~Niemiec:
    software.\\
N.~\v{S}ar\v{c}evi\'{c}:
    DESC Builder.\\
T.~Shin:
    methodology development;
    software;
    writing - original draft.\\
R.~Solomon:
    methodology development;
    software;
    writing - review \& editing.\\
R.~Srinivasan:
    methodology development;
    software;
    writing - original draft.\\
T.~Sunayama:
    writing - review \& editing.\\
Y.~Zhang:
    writing - review \& editing.\\
J.~Zuntz:
    writing - review \& editing

%% file: desc-tex/ack/standard.tex
The DESC acknowledges ongoing support from the Institut National de Physique Nucl\'eaire et de Physique des Particules in France; the  Science \& Technology Facilities Council in the United Kingdom; and the Department of Energy and the LSST Discovery Alliance in the United States.  DESC uses resources of the IN2P3 
Computing Center (CC-IN2P3--Lyon/Villeurbanne - France) funded by the  Centre National de la Recherche Scientifique; the National Energy  Research Scientific Computing Center, a DOE Office of Science User  Facility supported by the Office of Science of the U.S.\ Department of Energy under Contract No.\ DE-AC02-05CH11231; STFC DiRAC HPC Facilities,  funded by UK BEIS National E infrastructure capital grants; and the UK  particle physics grid, supported by the GridPP Collaboration. This  work was performed in part under DOE Contract DE-AC02-76SF00515.